# Roles of Individual Pigments in Ultrafast Excitation Dynamics of Light-Harvesting Phycobiliproteins Revealed by Recombinant Techniques and Two-Dimensional Electronic Spectroscopy

*Masaaki Tsubouchi,*[1,2,]* *Takatoshi Fujita,*[2] *Motoyasu Adachi,*[2] *and Ryuji Itakura*[1]

[1]Kansai Institute for Photon Science, National Institutes for Quantum Science and Technology (QST), 8-1-7 Umemidai, Kizugawa, Kyoto 619-0215, Japan

[2]Institute for Quantum Life Science, National Institutes for Quantum Science and Technology (QST), 4-9-1 Anagawa, Inage, Chiba 263-8555, Japan

**ABSTRACT:**

Phycobiliproteins serve as highly efficient light-harvesting antennae in cyanobacteria, yet the molecular factors governing their ultrafast energy relaxation and coherence dynamics remain incompletely understood. In this study, we investigate the role of pigment arrangement and pigment–protein interactions by combining recombinant protein engineering with two-dimensional electronic spectroscopy (2D-ES). In addition to wild-type allophycocyanin (APC) and C-phycocyanin (CPC), we artificially synthesized a β153 phycocyanobilin (PCB)–deficient CPC mutant, enabling direct experimental isolation of the contribution of this isolated pigment. The absorption and fluorescence spectra show that removal of the β153 pigment primarily eliminates its spectral contribution without significantly altering the excitonic coupling between the α84 and β84 pigments. Time-resolved 2D-ES reveals the close similarity between the dynamics of wild-type and β153-deficient CPCs, which demonstrates that the β153 pigment plays a minor role in ultrafast relaxation dynamics. Instead, the differences between APC and CPC arise primarily from pigment–protein interactions that modulate pigment structure and vibronic states. These results highlight the critical importance of local protein environments in controlling energy relaxation and coherence in photosynthetic light-harvesting proteins.

## ■ INTRODUCTION

Phycobilisomes are large, highly organized light-harvesting protein complexes found in cyanobacteria and red algae, enabling efficient light harvesting in diverse aquatic environments.[1] Energy transfer in phycobilisomes is known to be unidirectional with an extremely high quantum efficiency close to unity.[2] Recent advances in cryo-electron microscopy (cryo-EM) have provided high-resolution structural insights into phycobilisomes, greatly enhancing our understanding of their architecture and energy transfer mechanisms.[3-6] Complementary studies using ultrafast spectroscopies have revealed details of energy transfer and quantum coherence after photoexcitation, building on structural knowledge obtained from cryo-EM.[7-10] The phycobilisome complex is composed of multiple chromoproteins, primarily allophycocyanin (APC) and C-phycocyanin (CPC). Its core consists of APC assemblies, surrounded by CPC rods, with linker proteins connecting these components. Light energy absorbed by the phycobilisome is funneled toward the photosystem via the APC core. Within this energy transfer pathway, CPC and APC act as energy donors and acceptors, respectively.

Figures 1(a) and (b) show the structure and pigment arrangement of the APC trimer and CPC trimer, respectively, used in this study. Each APC monomer comprises α and β subunits, which contain phycocyanobilin (PCB) pigments covalently bonded to cysteine residues at positions α84 and β84, respectively. In the APC trimer, the α84 PCB and β84 PCB pigments from adjacent monomers are positioned in close proximity at a distance of around 2.1 nm.[11] In contrast, the CPC monomer contains a β153 PCB pigment in addition to α84 PCB and β84 PCB pigments. When a neighboring pair of α84 PCB and β84 PCB pigments is photoexcited, excitonic states can form through excitonic coupling between their electronic excited states. The β153 PCB pigment, however, remains decoupled from this pair, and its energy level of the electronic excited state is estimated to be higher than those of the excitonic states.[12] Delocalization of the electronic wavefunction across both α84 PCB and β84 PCB, driven by excitonic coupling, is thought to facilitate the unidirectional flow of energy, which has been extensively examined through both experimental and theoretical studies.[12-27]

As shown in Figs. 1(a) and (b), the primary structural difference between APC and CPC is the absence or the presence of the β153 PCB pigment. Although the β153 PCB pigment is electronically decoupled from the α84 PCB and β84 PCB pair, the photoexcitation dynamics differ remarkably between APC and CPC.[27] Investigating the molecular structure and spatial arrangement of pigments underlying this difference is crucial for understanding why nature has evolved phycobilisomes in their present form. Such insight could also guide the artificial design of antenna complexes with properties that surpass those found in nature.

To date, only wild-type APC and CPC isolated from native phycobilisomes have been used in spectroscopic investigations. However, this approach alone has made it difficult to directly interpret experimental results in terms of how the spatial arrangement of pigments and surrounding protein scaffolds influence the photoexcitation dynamics of the proteins. The aim of this study is to directly investigate the effect of pigment composition on the dynamics following photoexcitation by artificially synthesizing a β153 PCB pigment-deficient CPC shown in Fig. 1(c), thereby enabling a detailed examination of how pigment spatial arrangement influences the photoexcitation dynamics.

In recent years, the artificial synthesis of APC and CPC trimers, as well as their pigment-deficient variants, using *Escherichia coli* (*E. coli*) expression systems has been reported.[23, 28] We have also reported methods for the artificial synthesis of APC and CPC trimers in previous papers.[27] In this study, we further succeeded in the artificial synthesis of the β153

PCB pigment-deficient mutant CPC as shown in Fig. 1(c) and carried out spectroscopic measurements of its photoexcitation dynamics. The number of pigments in the mutant CPC is identical to that in the wild-type APC, and their spatial arrangements are also similar. In previous studies, the contribution of the β153 pigment was inferred from decomposition of the absorption spectra into multiple components. Here, we directly investigate the contribution of the β153 pigment by measuring the spectra of the β153 PCB pigment-deficient CPC in which the contribution of this pigment is eliminated. It is noted that the mutant CPC is not intended as a structural replica of the protein scaffold in APC, but rather as a system in which the peripheral pigment contribution is selectively removed while retaining that in CPC.

In this paper, we first briefly describe methods used to artificially synthesize the APC and CPC trimers that constitute the phycobilisome of *Thermosynechococcus elongatus* BP-1 and to determine their molecular structures. Additionally, we synthesized a mutant CPC lacking the β153 PCB pigment to compare its photoexcitation dynamics with those of the wild-type APC and CPC trimers. It has been suggested that the β153 PCB pigment makes a significant contribution to the static absorption spectrum, while playing a minor role in the dynamics of CPC. A comparison of the spectroscopic data between the wild-type and mutant CPCs allows the function of the β153 PCB pigment to be clarified experimentally.

To measure the ultrafast photoexcitation dynamics of synthesized proteins, we employ two-dimensional electronic spectroscopy (2D-ES).[27, 29] The 2D-ES has been developed to simultaneously probe energy relaxation kinetics and coherent dynamics resulting from photoexcitation to various vibronic states of light-harvesting proteins by ultrashort laser pulses with a broadband spectrum.[30-41] Despite its potential, the analysis has been challenging because multiple contributions are intricately mixed within the three-dimensional dataset measured by 2D-ES, $S(E_{\text{exc}}, E_{\text{det}}, T)$, where $E_{\text{exc}}$, $E_{\text{det}}$ and $T$ are the excitation and detection energies, and the delay time after photoexcitation, respectively. We introduce a method to derive kinetics components and coherent dynamics of proteins from the dataset $S(E_{\text{exc}}, E_{\text{det}}, T)$. Finally, we discuss the photoexcitation dynamics of APC and CPC and conclude with perspectives for future research.

## ■ METHOD

### Sample preparation

The light-harvesting proteins APC and CPC were designed on the basis of the structure determined by cryo-EM[6] and were independently produced by an *E. coli* expression system. APC and CPC proteins typically assemble into hexameric structures of heterodimers of α- and β-chains within the phycobilisome supercomplex. To facilitate spectroscopic and biochemical experiments, we introduced mutations at the intertrimeric interface to ensure trimer formation. This point is the primary difference from our previous study.[27]

Recently, Zhuang et al. reported the artificial synthesis of the mutant CPC trimers while controlling the number of pigments covalently bonded to them.[28] We have independently developed the artificial synthesis of the APC and CPC trimers and evaluated the structures of these phycobiliproteins by X-ray crystallography.[27] The recombinant *E. coli* strains and plasmids used in this study are presented in Fig. 1(d) and Supplementary Table S1, respectively. Two plasmids, pACYC_HO1PcyA and pCDF_MTSEF, were used for coexpression to enable pigment synthesis and protein modification, respectively. The culture conditions for recombinant *E. coli* and the protein purification were identical to those previously described.[27] Specifically, Gly21 and Gly29 in the α-chains of APC and CPC were substituted with arginine residues to introduce steric hinderance in hexamer formation, respectively. Furthermore, to match the pigment configuration of CPC to that of APC, we generated a mutant CPC lacking the β153 PCB pigment by substituting Cys153 with a tyrosine residue. Although substitution of the cysteine residue with any other amino acid results in loss of PCB, the tyrosine residue is thought to partially occupy the space left by the absence of PCB in its surrounding environment. In the following, unless otherwise specified, APC and CPC refer to the APC trimer and the CPC trimer, respectively. The mutant CPC refers to the PCB-deficient CPC at the β153 site.

To confirm the polypeptides present in the purified proteins, APC, CPC, and mutant CPC were analyzed by sodium dodecyl sulfate–polyacrylamide gel electrophoresis (SDS-PAGE). Figure 1(e) shows a photo of a stained gel. Two polypeptide bands corresponding to ApcA and ApcB appeared as a single band due to their similar migrations, while distinct bands for CpcA and CpcB were clearly observed. To further verify the presence of polypeptides and assess post-translational modification, intact protein mass spectrometry was performed. The measured masses for pigment-bound polypeptides were as follows: 18093.1 Da for ApcA and 19,822.0 Da for ApcB in wild-type APC, 18,158.0 Da for CpcA and 21,436.7 Da for CpcB in wild-type CPC, and 18,158.0 Da for CpcA and 20,910.1 Da for CpcB in mutant CPC. This analysis indicates that PCB is covalently attached to both the α- and β-chains, consistent with natural proteins purified from cyanobacteria. Furthermore, the β153 PCB is absent in the mutant CPC. Additionally, their molecular masses are consistent with the design that CpcM catalyzes methylation of asparagine residues in the β-chain. To examine the molecular assembly, the purified proteins, APC, CPC, and mutant CPC were analyzed by size-exclusion column chromatography (Fig. 1(f)). The major peaks for all three proteins appeared slightly after the 134 kDa marker, corresponding to the expected trimeric structure (approximately 110 kDa) formed by heterodimers of the α- and β-chains. These results indicate that all three proteins form trimers with the designed post-translational modifications.

For the 2D-ES measurement, proteins were diluted in buffer containing 10% (w/v) glycerol and 10 mM NaCl to achieve a concentration of approximately 40 μM (4.0 to 4.5 mg/mL). The solution was circulated through a sample cell with 1 mm-thick quartz windows and an optical path length of 0.2 mm using a peristaltic pump. The optical density of the absorption peak was approximately 0.4.

## Two-dimensional electronic spectroscopy

Absorption and fluorescence spectra were measured at room temperature (23 °C) by a JASCO spectrophotometer V-760 and a JASCO spectrofluorometer FP-8300, respectively.

The two-dimensional electronic (2D-E) spectra were measured using a previously reported setup.[27] The pump-probe configuration was employed in this study. An advantage of this configuration is that absorptive 2D-E spectra that directly visualize transient behavior can be obtained automatically with extremely high phase stability. In brief, we produced sub-10-femtosecond visible pulses using a Yb:KGW laser followed by two pulse-compression stages.[29] The coherent time interval between two pump pulses was scanned by the Translating-Wedge-Based Identical Pulses eNcoding System (TWINS)[42, 43] from −60 to 60 fs in steps of 0.4 fs. The waiting time, *T*, was scanned from 0 to 2000 fs in steps of 4 fs. The energy density of the excitation pulse irradiating the sample was 40 μJ/cm$^2$ in each pulse, which did not induce a nonlinear response by a single pulse. At the sample position, the linear polarizations of the two laser pulses were set in parallel. The duration of the excitation pulses was 13 fs at the sample position. The spectral width was sufficient to excite the entire absorption spectra of APC and CPC. One measurement, which collected about 500 2D-E spectra, was completed within three hours, and so multiple measurements with several proteins could be performed on the same day, indicating that quantitative comparison between different proteins is possible.

## ■ RESULTS

### Absorption and fluorescence spectra

The absorption spectra of the light-harvesting proteins synthesized in this study are shown in Fig. 2(a) as solid lines. The intensities are normalized at a wavelength of 278 nm, corresponding to the peak assigned to cysteine residues present in all three proteins in equal amounts. The spectrum of APC exhibits a peak and a shoulder at around 650 nm and 600 nm, which have been assigned to the lower and higher excitonic states, respectively.[15] On the other hand, the spectrum of CPC does not clearly exhibit a shoulder, but it has a tail on the short-wavelength side. The peak wavelength is 620 nm, which is 30 nm shorter than that of APC. In the mutant CPC, the peak wavelength is located at around 625 nm, which is close to that of the wild-type CPC. However, the spectral shape exhibits a main peak accompanied by a shoulder, resembling that of APC. To quantitatively compare the absorption spectra, the spectra are decomposed into three Gaussian components using nonlinear least-squares fitting, as described in Supplementary Sec. 3. The peak wavelengths and the widths of three fitted components are summarized in Table 1.

Because the mutant CPC lacks the β153 pigment and contains only two types of pigments, the two-component Gaussian fit would be expected to be sufficient to describe the absorption spectrum. However, three components are required to adequately reproduce the spectrum. The peak positions and spectral widths of the three components in the mutant CPC are nearly identical to those of the wild-type CPC. Although the mutant CPC spectrum appears to exhibit a shoulder, the experimental data can be well reproduced simply by changing the relative intensities of the three components compared to those of the wild type. With regard to the contribution of the β153 pigment, it may be necessary to introduce a fourth component in the fitting for the wild-type spectrum. However, distinguishing between the fitting results obtained with three and four components is difficult based on the experimental spectra alone.

The wild-type APC also contains only two types of pigments. However, its absorption spectrum cannot be adequately fitted with two Gaussian functions and instead requires three Gaussian components, similar to the mutant CPC. Womick and Moran suggested that vibronic coupling plays an important role in the energy relaxation dynamics of phycobiliproteins.[20] Figure 2(b) shows the energy diagram of the excitonic states when the vibronic coupling is taken into account. The lowest-energy peak (peak A in Table 1) may correspond to the ground vibrational state of the α84 pigment. The peaks B and C may correspond to the lower and higher excitonic states, respectively, which are composed of the excited vibrational state of the α84 pigment and the ground vibrational state of the β84 pigment.

A comparison between the wild-type APC and CPC reveals that the energy difference between the peaks are nearly identical as shown in Table 1, with the only notable difference being that the absolute values of the site energies are higher in APC than in CPC. The excitonic couplings in APC are therefore expected to be similar to those in CPC, which is inconsistent with the conclusion of previous studies.[12, 18-20]

The absence of the β153 pigment in the mutant CPC does not result in a decisive change in the spectral shape and has little effect on excitonic coupling between the α84 and β84 pigments. The difference in site energies between APC and CPC is most likely attributable to differences in

pigment–protein interactions arising from variations in the surrounding amino-acid framework as explained later.

The fluorescence spectra of the three phycobiliproteins are shown as dashed lines in Fig. 2(a). As described in Supplementary Sec. 3, these spectra are well reproduced by the sum of two Gaussian functions. The peak wavelengths and spectral widths obtained from the fits are summarized in Table 1. The appearance of two components in the fluorescence spectra can be attributed to spontaneous emission from pigments located in two distinct local environments.

The most notable feature is that the fluorescence spectrum of the mutant CPC is almost identical to that of the wild-type CPC except for the tail on the short-wavelength side. This result indicates that the photon energy absorbed by the β153 pigments in CPC is transferred to the excitonic states of the α84 and β84 pigment pairs and subsequently relaxes to the bottom of the potential energy surface of the lowest electronic excited state prior to spontaneous emission. Furthermore, the close similarity between the fluorescence spectra of the wild-type and mutant CPCs suggests that the potential energy surface associated with the excitonic states of the α84-β84 pigment pair is scarcely affected by the presence of the β153 pigment in CPC. In addition, the fluorescence spectrum of CPC exhibits an asymmetric profile with a broadened tail on the long-wavelength side, whereas that of APC shows a symmetric shape about the peak wavelength.

Thus, based on the results of static absorption and fluorescence spectroscopy, we conclude that even when the β153 PCB pigment is removed to make the pigment arrangement of CPC similar to that of APC, the fundamental characteristics of the excitonic states do not approach those of APC. Rather, the primary consequence is the disappearance of the spectral contribution from the β153 pigment, which is hardly coupled to the α84 and β84 pigment pair. The contribution of the β153 PCB pigment to the fluorescence spectrum of CPC was previously discussed by Womick and Moran.[12] In the present study, we artificially synthesized a β153-deficient CPC mutant and, by comparing it with the wild-type CPC, experimentally clarified the role of the β153 pigment for the first time.

## Two-dimensional electronic spectra

Figure 3 presents a comparison of time-resolved 2D-E spectra, $S(E_{\mathrm{exc}}, E_{\mathrm{det}}; T)$, of APC, CPC, and the mutant CPC. The horizontal and vertical axes show the excitation ($E_{\mathrm{exc}}$) and detection ($E_{\mathrm{det}}$) energies, respectively. Positive amplitudes indicate increased transmission, corresponding to the ground state bleaching (GSB) or the stimulated emission (SE), while negative amplitudes indicate the excited state absorption (ESA). Immediately after excitation ($T$ = 0 fs), the signal lies along the diagonal, indicating the absence of energy relaxation on the time scale of the pulse duration.

At $T$ = 1 ps, the amplitude distribution in the 2D-E spectrum of APC becomes aligned parallel to the horizontal axis at $E_{\mathrm{det}}$ = 15,240 cm$^{-1}$, which is close to the peak of the fluorescence spectrum of APC (15,210 cm$^{-1}$) shown in Fig. 2(a). This suggests that all the excited vibronic states populated by the pump pulses have nearly completely relaxed to the bottom of the potential energy surface of the lowest electronic excited states within 1 ps. On the other hand, in the 2D-E spectra of CPC and its mutant, the positive signals spanning a broad detection-energy range higher than 15,270 cm$^{-1}$ are detected at $T$ = 1 ps, unlike in the case of APC. This result indicates that relaxation to the bottom of the potential is not completed and, therefore, the excited vibronic states still survive at a few picosecond time delays after the excitation. The ESA component also appears at a detection energy below $E_{\mathrm{det}}$ = 15,270 cm$^{-1}$, which is not clearly seen in the case of APC. Since the ESA component is not present immediately after excitation, it must originate from a process involving states populated after energy relaxation.

The 2D-E spectra of the mutant CPC lacking the β153 pigments more closely resemble those of the wild-type CPC than the wild-type APC. This tendency is consistent with that observed in the absorption and fluorescence spectra. For a quantitative comparison of the 2D-E spectra, Fig. 4 presents the transient spectra of three phycobiliproteins at a long delay time ($T$ = 2 ps), derived by integrating the 2D-E spectra along the excitation energy axis. The spectrum of the mutant CPC differs remarkably from that of APC, while being nearly identical to that of the wild-type CPC, except for a red shift of approximately 40 $cm^{-1}$ and a weak dip around 16,500 $cm^{-1}$. This weak dip is attributed to the absence of transient species associated with the β153 pigment, and its energy is close to that of peak B in the absorption spectra of CPCs.

## Transient behavior of lowest electronic excited states

A series of 2D-E spectra $S(E_{\mathrm{exc}}, E_{\mathrm{det}}; T)$ is reconfigured into a series of transient spectra obtained by slicing the three-dimensional data set of $S(E_{\mathrm{exc}}, E_{\mathrm{det}}, T)$ at specific excitation energies. Figure 5(a) shows the transient spectrum, $S(E_{\mathrm{det}}, T; E_{\mathrm{exc}})$, of APC at $E_{\mathrm{exc}}$ = 15,550 $cm^{-1}$, which is 210 $cm^{-1}$ higher than the energy of the absorption peak A corresponding to the lowest electronic excited state. The transient spectrum exhibits not only a dynamic Stokes shift but also strong oscillatory behavior. The dynamic Stokes shift rapidly completes within 500 fs, and the detection energy converges to $E_{\mathrm{det}}$ = 15,250 $cm^{-1}$, which is close to the peak energy in the fluorescence spectrum.

To understand the kinetics of the photoexcited molecules, Global analysis based on singular value decomposition has been widely applied to the analysis of transient spectra. However, when coherent features overlap with kinetic components, the experimental data cannot be accurately reconstructed using a small number of principal components obtained through singular value decomposition.[44, 45] Therefore, the time profiles obtained at all pairs of excitation and detection energies, $S(T;\ E_{\mathrm{exc}}, E_{\mathrm{det}})$, are fitted with a single exponential function,

$$S(T;\ E_{\mathrm{exc}}, E_{\mathrm{det}}) = A_0 + A_1 \exp(-T/\tau_1), \tag{1}$$

where $\tau_1$ is the time constant of the exponential decay, and $A_0$ and $A_1$ are the amplitudes of the constant and the decay components, respectively. These three parameters are functions of $E_{\mathrm{exc}}$ and $E_{\mathrm{det}}$. The kinetic component and coherent dynamics are then separated based on the fitted profiles and their residuals, respectively. In this study, we focus on dynamics occurring within 2 ps after photoexcitation. Therefore, the energy relaxation process is described by a single-exponential function, while all processes that evolve sufficiently slowly on this time scale are incorporated into a constant term. Further details are provided in Supplementary Sec. 3.

Figures 5(b) and (c) display the kinetic and residual (coherent) components, respectively, obtained by decomposing the transient spectrum measured at $E_{\mathrm{exc}}$ = 15,550 $\mathrm{cm}^{-1}$ shown in Fig. 5(a). In the coherent component, the pronounced beating feature is evident, exhibiting a node around the detection energy of $E_{\mathrm{det}}$ = 15,250 $\mathrm{cm}^{-1}$. The sign of the amplitude reverses at the node, corresponding to a $\pi$ phase shift in the beating pattern. This provides clear evidence of vibrational wave packet motion on the electronic excited state, as discussed in our previous study.[27]

Figures 5(d) and (g) present the transient spectra $S(E_{\mathrm{det}}, T; E_{\mathrm{exc}})$ of the wild-type and mutant CPC obtained at the excitation energies $E_{\mathrm{exc}}$ = 16,130 $\mathrm{cm}^{-1}$ and 16,200 $\mathrm{cm}^{-1}$, respectively. These excitation energies are about 200 $\mathrm{cm}^{-1}$ higher than those of the absorption peak A. In the transient spectra of both CPCs, the GSB or SE component with positive amplitude overlaps with the negative ESA component around 15,300 $cm^{-1}$. Figures 5(e) and (h) show the kinetic component of the wild-type and the mutant CPCs, respectively. The dynamic Stokes shift converges within 100 fs, which is much faster than that observed in APC. In addition, broad transient spectral features appear immediately after photoexcitation and persist throughout the observation window. This behavior is consistent with rapid vibrational energy redistribution over a wide manifold of vibronic states. However, we note that the broader absorption linewidth of CPC relative to APC may also contribute to the observed spectral broadening. Regardless of its microscopic origin, the transient spectrum, even at 2 ps, remains substantially broader than the fluorescence spectrum as shown in Fig. 4, indicating that complete relaxation to the bottom of the potential energy surface is not achieved within the present observation window. This behavior is distinctly different from that of APC, in which the vibrational energy almost completely relaxes on the same time scale. The transient spectrum of wild-type APC at $T$ = 2 ps is nearly identical to the fluorescence spectrum, which is consistent with our earlier conclusion that vibrational relaxation is completed within 2 ps in APC.

Figures 5(f) and (i) show the coherent component of the wild-type and the mutant CPCs, respectively. The beating signals decay rapidly within a few hundred femtoseconds, and the nodal line is not discernible. This behavior is also markedly different from that of APC.

## Transient behavior of higher excitonic states

Figure 6(a) shows the transient spectrum of APC at $E_{\mathrm{exc}}$ = 16,670 $cm^{-1}$, which is close to the energy of the absorption peak C corresponding to the higher excitonic state. The energy relaxation occurs rapidly within several hundred femtoseconds, and the detection energy converges to $E_{\mathrm{det}}$ = 15,250 $cm^{-1}$, which is similar to the case in which the lowest electronic excited state is excited. As shown in Fig. 6(c), beat signals are barely observed except for weak residual signals around $E_{\mathrm{det}}$ = 15,250 $\mathrm{cm}^{-1}$. However, these residual signals cannot be reproducibly observed across multiple experiments within the signal-to-noise ratio limitation of our apparatus.

The transient spectra of the wild-type and the mutant CPCs are shown in Figs. 6(d) and (g), respectively. These spectra are obtained at the excitation energies of $E_{\mathrm{exc}}$ = 17,210 $\mathrm{cm}^{-1}$ and 17,140 $\mathrm{cm}^{-1}$ corresponding to the energy of the absorption peak C of the wild-type and the mutant CPCs, respectively. As in the case of excitation to the lowest electronic excited states, a rapid dynamic Stokes shift occurs within 100 fs, accompanied by vibrational energy redistribution. The resulting energy broadening is also nearly identical to that observed upon excitation to the lowest electronic excited states. These results indicate that the dynamics following the rapid internal conversion are essentially the same regardless of which electronic state is initially excited.

Here, we summarize the dynamics presented in this section. In APC, internal conversion occurs rapidly within 100 fs, followed by vibrational energy relaxation on a time scale of approximately 500 fs, driving the system to the bottom of the potential energy surface. In contrast, in both wild-type and mutant CPCs, internal conversion similarly takes place within 100 fs, after which the excitation energy is redistributed over a broad range of vibronic states. On a time scale of about 2 ps, no further vibrational energy relaxation is observed. The β153 pigment in CPC does not play a dominant role in the energy relaxation dynamics. Rather, the differences in relaxation behavior between APC and CPC are primarily governed by the interaction between the pigments and the amino-acid scaffold. The artificial synthesis of the pigment-deficient mutant CPC makes it possible to clearly discuss these aspects.

## Quantitative analysis of energy relaxation

When decomposing the transient spectra into kinetic and coherent components as shown in Figs. 5 and 6, the time profiles, $S(T;\ E_{\mathrm{exc}}, E_{\mathrm{det}})$, at all grid points of excitation and detection energies are fitted using Eq. (1), and the parameters, $A_0$, $A_1$, and $\tau_1$, are determined. Figure 7 presents two-dimensional maps of the parameters $A_0$, $A_1$, and $\tau_1$, for three phycobiliproteins. As expected, the map of the constant component $A_0$ (Figs. 7(a), (d), and (g)) closely resembles the 2D-E spectrum measured at a long delay time ($T$ = 1 ps) shown in Figs. 3(c), (f), and (i). Figure 7(b) shows the map of the parameter $A_1$ for APC corresponding to the two-dimensional decay-associated spectra (2D-DAS). At the detection energies higher and lower than $E_{\mathrm{det}}$ ~ 15,400 cm$^{-1}$, the amplitude of $A_1$ is positive and negative, respectively. A positive $A_1$ indicates the population decays, whereas a negative $A_1$ indicates population growth. The boundary is located around $E_{\mathrm{det}}$ = 15,400 cm$^{-1}$, which is close to the peak of the absorption spectrum of APC. Black and white dots appear in the 2D maps, particularly around the detection energy of $E_{\mathrm{det}}$ ~ 15,400 cm$^{-1}$. These indicate that the fitting does not converge well, because the observed time profiles lack any noticeable decaying or rising features.

The 2D-map of the parameter $A_1$ (hereafter referred to as the $A_1$ map) for wild-type and mutant CPC are shown in Figs. 7(e) and 7(h), respectively. The boundary between positive and negative signals appears at $E_{\mathrm{det}}$ ~ 15,300 cm$^{-1}$, which seems to be similar to that observed for APC. However, the boundary in CPCs almost coincides with the energy boundary observed in the $A_0$ maps, which is different from the case of APC, where the boundary in the $A_1$ map coincides with the ridge of the $A_0$ map. The positive component in the $A_1$ map reflects the decay of GSB or SE with positive amplitude in the 2D-E spectra, whereas the negative component corresponds to the decay (rather than the rise) of ESA with negative amplitude in the 2D-E spectra. Although rise components associated with population transfer to lower vibronic states should in principle also be present, they are likely obscured by the dominant ESA contributions and thus remain unresolved.

The 2D-map of the decay time-constant $\tau_1$ for APC shown in Fig. 7(c) consists of two components. One is a fast decay component ($\tau_1$ < 200 fs) along the diagonal. This component likely reflects ultrafast relaxation processes, which may include internal conversion and/or rapid structural relaxation. Because it is difficult to clearly distinguish between these processes in our experiment, we here tentatively assign this rapid relaxation process as internal conversion. The other is a relatively slow decay or rise component ($\tau_1$ ~ 500 fs) located below the diagonal. These findings suggest dynamics of photoexcited APC involving fast internal conversion, followed by vibrational energy relaxation to the bottom of the potential energy surface of the electronic excited state.[13-15, 18]

To enable a more quantitative discussion, the 2D-map is divided into several regions based on the amplitudes in the 2D-DAS. The distributions of the time constants $\tau_1$ obtained at the individual ($E_{\mathrm{exc}}, E_{\mathrm{det}}$) grid points within each specific region are then presented as histograms.[46] Figures 8(a) and 8(b) show the $A_1$ maps for APC in which only the positive and negative components are highlighted, respectively. Regions with very small amplitudes indicate negligible temporal variation in the time profiles, for which the fitting does not reliably converge. Therefore, the lower threshold on the absolute amplitude, 0.01, is introduced to exclude regions exhibiting little population decay or rise.

Histograms of the time constants $\tau_1$ for the regions of the positive and negative amplitudes in the 2D-DAS of APC are shown in the upper and lower panels of Fig. 8(c), respectively. The decay time constant derived from the positive regions is $\tau_1 = 420 \pm 150$ fs, while the rise time constant obtained from the negative regions is $\tau_1 = 370 \pm 110$ fs, indicating that vibrational energy relaxation to the bottom of the potential energy surface occurs on a timescale of approximately 400 fs. Closer inspection of the decay components reveals additional contributions with time constants shorter than 200 fs. To analyze these fast components in more detail, the time constants within only the region bounded by the dashed lines near the diagonal in Fig. 8(a) are extracted and compiled into a histogram, as shown in the middle panel of Fig. 8(c). In this region, the fast decay time constants are clearly observed, revealing two components with $\tau_1 = 110 \pm 80$ fs and $280 \pm 57$ fs. As seen in Fig. 7(c), the faster component appears at detection energies corresponding to absorption peaks B (15,830 $cm^{-1}$) and C (16,650 $cm^{-1}$), whereas the slightly slower component is observed primarily in the intermediate energy region between these peaks. This result indicates that in APC, the initially populated higher excitonic states undergo internal conversion on a time scale of approximately 100 fs immediately after photoexcitation, followed by completion of vibrational relaxation over roughly 400 fs.

The sub-picosecond ultrafast dynamics of APC have been explored by measuring and calculating energy relaxation and quantum coherence.[13-15, 18] The fastest step is the interexcitonic-state relaxation with the vibrational energy redistribution at a time constant of 30 fs during dephasing of the electronic and nuclear coherence. The change in geometric configurations of the pair of pigments occurs within a time constant of 200–300 fs after the first step mentioned above, followed by the localization of energy. This process has been considered as a key to unidirectional energy transfer. The decay components identified in the present study are quantitatively consistent with these earlier observations.

In contrast, as described above, only decay components are observed for CPC. As shown in Figs. 8(d) and 8(e), the regions in the 2D-DAS that exhibit significant decay within 2 ps are confined to areas near the diagonal. Figure 8(f) shows histograms of the decay time constants $\tau_1$ for wild-type and mutant CPCs. When the analysis is restricted to the region between the dashed lines, as is performed for APC, the histograms reveal that both wild-type and mutant CPC exhibit a peak around 20 fs. This indicates that CPC undergoes much faster internal conversion as compared to APC, after which vibrational energy redistribution populates a broad range of vibronic states. The subsequent vibrational energy relaxation requires much more than 2 ps.

Previous spectroscopic investigations suggested that the β153 pigment contributes significantly to the absorption spectrum of CPC while participating only weakly in the excitonic states formed by the α84 and β84 pigments.[12, 21] The present measurements directly confirm this picture by showing that removal of the β153 pigment produces only minor changes in the ultrafast relaxation and coherence dynamics despite noticeable changes in the absorption spectrum. The new aspect of the present work is the direct comparison among APC, wild-type CPC, and a pigment-deficient CPC mutant under identical experimental conditions, which enables the contribution of the β153 pigment to be experimentally isolated.

## Vibrational coherence

The coherent nature of the excitonic states is also one of the central concerns of photoexcited dynamics of phycobiliproteins. The detailed method for the analysis of coherence dynamics is provided in Supplementary Sec. 5. The beat frequency spectrum of APC obtained at the excitation energy $E_{\mathrm{exc}}$ = 15,550 cm$^{-1}$ is shown in the upper panel of Fig. 9(a). We found that the peak beat frequency is $\nu_{\mathrm{beat}}$ = 202 cm$^{-1}$ and the full width at half maximum is 46 cm$^{-1}$. The broad width of the peak indicates a short decoherence time on the order of a few hundred femtoseconds. The beat spectra were calculated for all pairs of excitation and detection energies $(E_{\mathrm{exc}}, E_{\mathrm{det}})$, and the intensities at $\nu_{\mathrm{beat}}$ = 202 cm$^{-1}$ in the beat spectra are plotted as a two-dimensional map, as shown in Fig. 9(b). The two peaks are observed with a nodal line running parallel to the excitation energy axis around the detection energy of $E_{\mathrm{det}}$ = 15,250 cm$^{-1}$. In Fig. 9(b), the contour map of the 2D-E spectrum observed at $T$ = 1 ps is overlaid on the beat intensity map. The nodal line of the beat intensity map traces the ridge of the 2D-E spectrum, corresponding to the bottom of the potential energy surface of the lowest electronic excited states. Figure 9(c) presents the spectra of the beat intensity and phase at an excitation energy of $E_{\mathrm{exc}}$ = 15,550 cm$^{-1}$ as a function of the detection energy. In the phase spectrum, a clear π-phase jump is observed at the detection energy of $E_{\mathrm{det}}$ = 15,250 cm$^{-1}$, corresponding to the nodal line in the 2D map. These results provide evidence for the formation of a vibrational wave packet on the electronic excited state, as illustrated in Fig. 9(d).[7] The components of A and B indicated in Figs. 9(b) and (c) correspond to the turning points on the potential energy surface. The decoherence time of the beating signal as a function of the detection energy is overlaid in the beat intensity spectrum in Fig. 9(c). Around the detection energy of $E_{\mathrm{det}}$ = 15,250 cm$^{-1}$, the fitting results of the decoherence time are not reliable, because the beat intensity is too weak to fit to the oscillating model function. The decoherence time in the region A is 140 ± 20 fs, on the other hand, that in the region B is 370 ± 80 fs. These decoherence times are similar to the time constant for the vibrational relaxation, $\tau_1 \sim 400$ fs.

In contrast, for CPC, the beat signals are not clearly discernible, as shown in Figs. 5(f) and 5(i). No distinct nodal lines are observed, and the decoherence occurs more rapidly than in APC. Figure 9(a) compares the beat spectra of three phycobiliproteins, in which a peak is observed around 200 $cm^{-1}$. The peak intensities of both CPCs are comparable to that of low-frequency noise below 50 $cm^{-1}$. This feature is independent of the presence or absence of the β153 pigment. Thus, CPC exhibits a coherent nature markedly different from that of APC. While this difference is likely related to variations in the amino-acid scaffold surrounding the pigments, the decisive factor responsible for the distinct behaviors of APC and CPC cannot be identified based on the present results and remains an open question for future studies.

In the previous studies, the vibrational mode promoting the energy relaxation has been suggested to be the hydrogen out-of-plane (HOOP) wagging motion with a frequency of 800 $cm^{-1}$.[18] This frequency is close to the site energy difference between pigments; therefore, it has been considered that the vibronic states with this promoting mode contribute to the delocalized excitonic states. Additionally, it is possible that the low-frequency mode observed in our beat-frequency spectra is also the mode promoting energy transfer. However, this mode has not been considered to be an important vibrational mode for the relaxation process because it is in-plane isotropic motion that barely induces geometrical change through the vibration. Sohoni et al. performed transient absorption measurements on intact phycobilisomes and analyzed the vibrational coherence of APC and CPC. Their results suggested that the vibrational modes promoting inter-excitonic energy transfer in CPC are lower-frequency modes rather than HOOP modes in contrast to APC, where HOOP modes were identified as the promoting vibrations.[26] As a result, the previous studies have not yet reached a consensus regarding the key low-frequency promoting modes. Because the molecular structure is an important factor in determining the energy structure of the vibronic states and the energy relaxation dynamics, the lower-frequency mode, $\nu_{\mathrm{beat}}$ = 202 cm$^{-1}$, is also expected to remain important. The low-frequency coherence observed at approximately 202 cm$^{-1}$ may originate from several possible sources. One possibility is an intramolecular deformation mode of the PCB pigment. Alternatively, the mode could reflect intermolecular pigment-protein vibrations or collective motions involving both the pigment and the surrounding protein matrix. At present, the available data do not permit a definitive assignment. However, the strong appearance of this coherence in APC and its absence in CPC suggest that the local protein environment plays an important role in determining the extent to which such motions remain coherent.

## ■ DISCUSSION

As explained in the previous section, the energy relaxation and coherence dynamics of the wild-type CPC differ significantly from those observed in the wild-type APC and, in contrast, are very similar to those of the mutant CPC. The results obtained in this study indicate that the spatial arrangement of pigments is not the only significant factor that determines the photo-induced dynamics of light-harvesting proteins. The effect of the absence of the pigment covalently bonded to the Cys153 is mainly the elimination of light absorption by the β153 PCB and the associated energy transfer from it to the exciton pair formed by α84 and β84 PCBs. These observations suggest that the primary biological role of the β153 pigment is to function as a peripheral light-harvesting antenna. The β153 pigment broadens the absorption range and increases the effective absorption cross-section of CPC while transferring the absorbed excitation energy to the α84–β84 excitonic states prior to fluorescence emission.

Figure 10 compares the stereo structures of the pigment PCBs in the wild-type APC and the wild-type CPC. The structures of β84 PCB in APC and CPC appear similar. In contrast, the α84 PCB in APC has a planar structure, whereas that in CPC is bent, as clearly seen in the side view of the PCB highlighted by the red circle. Although the specific amino-acid residues responsible for this structural difference cannot be identified from the present measurements, the difference in the local protein environments surrounding the α84 PCB pigments between APC and CPC may lead to variations in hydrogen-bonding interactions and steric constraints imposed by neighboring residues, thereby influencing chromophore planarity. This difference may be caused by a reduction in π-conjugation at the position indicated by the red circle in PCB. The structural change alters the electronic structure and, consequently, the photo-induced dynamics of APC compared to CPC. So far, investigations of energy transfer in phycobilisomes have focused on the distance and relative orientation between the pigment molecules.

The conclusion follows directly from the comparison among APC, wild-type CPC, and the β153-deficient CPC mutant. APC and CPC exhibit markedly different relaxation and coherence dynamics. However, removing the β153 pigment does not make mutant CPC behave more like APC. Instead, the mutant remains spectroscopically and dynamically similar to wild-type CPC. These results demonstrate that pigment composition and spatial arrangement alone are insufficient to explain the observed dynamics. The remaining differences must therefore arise primarily from local pigment-protein interaction that affects chromophore geometry, site energies, and vibronic coupling. These interactions determine the molecular structure of the pigments, which in turn governs their electronic structures and photoexcitation dynamics.[19] Therefore, understanding how structural variations influence the dynamics is essential for clarifying the role of pigment–protein interactions. For this purpose, further experiments (e.g., introducing mutations that alter the planarity of the pigment) and numerical calculations are required. Because AlphaFold-based predictions remain unreliable for ligand structures such as pigments, systematic repetition of mutant synthesis and crystallographic analysis are necessary.

The distinct relaxation pathways observed in APC and CPC may also be connected to their different physiological functions within the phycobilisome. APC is located in the core region and serves as the terminal emitter that transfers excitation energy toward the photosystems. Rapid relaxation toward the lowest-energy excited state may therefore help establish efficient directional energy flow. In contrast, CPC functions as an intermediate antenna protein in the peripheral rods. The broader distribution of vibronic states observed in CPC may facilitate acceptance of excitation energy over a wider energy range while maintaining efficient transfer toward the lower-energy APC core. Although the present experiments do not directly establish this functional relationship, the observed dynamics are qualitatively consistent with the distinct biological roles of these proteins.

## ■ CONCLUSIONS

We investigated the roles of pigment arrangement and pigment–protein interactions in governing ultrafast photoexcitation dynamics in light-harvesting phycobiliproteins. By artificially synthesizing a β153 PCB–deficient CPC mutant and comparing its spectroscopic properties with those of wild-type CPC and APC, we were able to directly isolate the contribution of the peripheral β153 pigment. Static absorption and fluorescence spectroscopy revealed that removal of the β153 pigment primarily eliminates its spectral contribution without substantially altering the excitonic coupling between the α84 and β84 pigments. 2D-ES further demonstrated that, although internal conversion occurs within ~100 fs in all proteins examined, the subsequent vibrational energy relaxation differs markedly between APC and CPC. In APC, excitation energy rapidly relaxes to the bottom of the potential energy surface within sub-picosecond time scales, accompanied by pronounced vibrational coherence. In contrast, both wild-type and β153-deficient CPCs exhibit rapid vibrational energy redistribution over a broad manifold of vibronic states, with no further relaxation observed within 2 ps and only weak, short-lived coherence signals.

The close similarity between the photoexcitation dynamics of wild-type and mutant CPCs clearly indicates that the β153 pigment does not play a dominant role in the ultrafast relaxation and coherence dynamics. Instead, the distinct behaviors of APC and CPC are primarily determined by differences in pigment–protein interactions that modulate pigment geometry, site energies, and vibronic coupling. These findings demonstrate that pigment spatial arrangement alone is insufficient to describe excitation dynamics, and local structural details of the protein scaffold are equally crucial. More generally, the present results establish an experimental framework for separating the influence of pigment composition from that of the local protein environment, thereby providing new insight into the molecular factors governing ultrafast excitation dynamics in photosynthetic antenna proteins.

By employing recombinant techniques to create a pigment-deficient mutant, this work provides a direct experimental approach to dissect the molecular origins of energy relaxation and coherence in photosynthetic antenna proteins. The results highlight the importance of pigment–protein interactions in natural light-harvesting systems and offer guiding principles for the rational design of artificial antenna complexes with tailored ultrafast dynamics.

To further validate the structural elements that govern kinetics and coherence in light-harvesting proteins, we plan to synthesize the various mutant APC and CPC proteins. For example, certain mutations are designed to prevent the formation of excitonic states or disrupt the energy balance within these states. These future investigations will highlight a critical role of local pigment-protein and pigment-pigment interactions in ultrafast energy relaxation and coherence, offering valuable insights for the design of artificial light-harvesting systems with tailored dynamics.

## ASSOCIATED CONTENT

**Supporting Information**.

The Supporting Information is available free of charge at xxxx.

Detailed information on sample preparation and the additional dataset (PDF).

## AUTHOR INFORMATION

**Corresponding Author**

**Masaaki Tsubouchi** – Kansai Institute for Photon Science, National Institutes for Quantum Science and Technology (QST), 8-1-7 Umemidai, Kizugawa, Kyoto 619-0215, Japan; https://orcid.org/0000-0003-1158-0867; Email: tsubouchi.masaaki@qst.go.jp

**Authors**

**Takatoshi Fujita** – Institute for Quantum Life Science, National Institutes for Quantum Science and Technology (QST), 4-9-1 Anagawa, Inage, Chiba 263-8555, Japan; https://orcid.org/0000-0003-1504-2249

**Motoyasu Adachi** – Institute for Quantum Life Science, National Institutes for Quantum Science and Technology (QST), 4-9-1 Anagawa, Inage, Chiba 263-8555, Japan; https://orcid.org/0000-0003-2353-880X

**Ryuji Itakura** – Kansai Institute for Photon Science, National Institutes for Quantum Science and Technology (QST), 8-1-7 Umemidai, Kizugawa, Kyoto 619-0215, Japan; https://orcid.org/0000-0003-0508-4760

**Notes**

The authors declare no competing financial interest.

**ACKNOWLEDGMENT**

We thank Dr. Yoshihisa Ishii, Yuji Kagotani and Takemi Kanazawa at QST for their assistances in constructing the laser system, in performing the measurements of protein absorption and fluorescence spectra, and in protein synthesis, respectively. We also thank Prof. A. Ishizaki at The University of Tokyo for valuable comments about 2D-ES. This research was supported by the MEXT Quantum Leap Flagship Program (JPMXS0120330644). We are grateful for the financial support of JSPS KAKENHI (JP21H01898) and the QST President's Strategic Grant (Exploratory Research). The mass spectrometry analysis shown in the Supplementary Materials was supported by J-PARC MLF deuteration laboratory.

**Table 1**. Spectroscopic data for APC, CPC, and PCB-deficient CPC at the β153 site (mutant CPC). The upper row in each cell shows the peak frequency, and the lower row shows the full width at half-maximum (FWHM) of each peak. These data were obtained by fitting the absorption and fluorescence spectra to the sum of three and two Gaussian functions, respectively, as described in Supplementary Section 3.

| | APC | CPC | Mutant CPC |
|---|---|---|---|
| Abs. peak A | 15,340 $cm^{-1}$ (652.0 nm) | 15,910 $cm^{-1}$ (628.5 nm) | 15,940 $cm^{-1}$ (627.6 nm) |
| FWHM | 450 $cm^{-1}$ | 650 $cm^{-1}$ | 700 $cm^{-1}$ |
| Abs. peak B | 15,830 $cm^{-1}$ (631.7 nm) | 16,410 $cm^{-1}$ (609.6 nm) | 16,390 $cm^{-1}$ (610.0 nm) |
| FWHM | 1120 $cm^{-1}$ | 1180 $cm^{-1}$ | 1200 $cm^{-1}$ |
| Abs. peak C | 16,650 $cm^{-1}$ (600.5 nm) | 17,210 $cm^{-1}$ (581.1 nm) | 17,120 $cm^{-1}$ (584.0 nm) |
| FWHM | 2160 $cm^{-1}$ | 2320 $cm^{-1}$ | 2190 $cm^{-1}$ |
| $\Delta E(B - A)$[a] | 490 $cm^{-1}$ | 500 $cm^{-1}$ | 450 $cm^{-1}$ |
| $\Delta E(C - B)$[b] | 850 $cm^{-1}$ | 800 $cm^{-1}$ | 730 $cm^{-1}$ |
| Fluor. Peak A | 15,210 $cm^{-1}$ (657.4 nm)[c] | 15,680 $cm^{-1}$ (637.9 nm) | 15,650 $cm^{-1}$ (639.0 nm) |
| FWHM | 490 $cm^{-1}$, 1330 $cm^{-1}$ | 720 $cm^{-1}$ | 690 $cm^{-1}$ |
| Fluor. Peak B | | 15,510 $cm^{-1}$ (644.9 nm) | 15,350 $cm^{-1}$ (651.5 nm) |
| FWHM | | 1600 $cm^{-1}$ | 1440 |
| Stokes shift[d] | 130 $cm^{-1}$ | 230 $cm^{-1}$ | 350 $cm^{-1}$ |

[a]Frequency difference between peaks A and B.

[b]Frequency difference between peaks B and C.

[c]Two Gaussian components that share the same peak position but have different widths.

[d]Frequency difference between the absorption peak A and the fluorescence peak A.

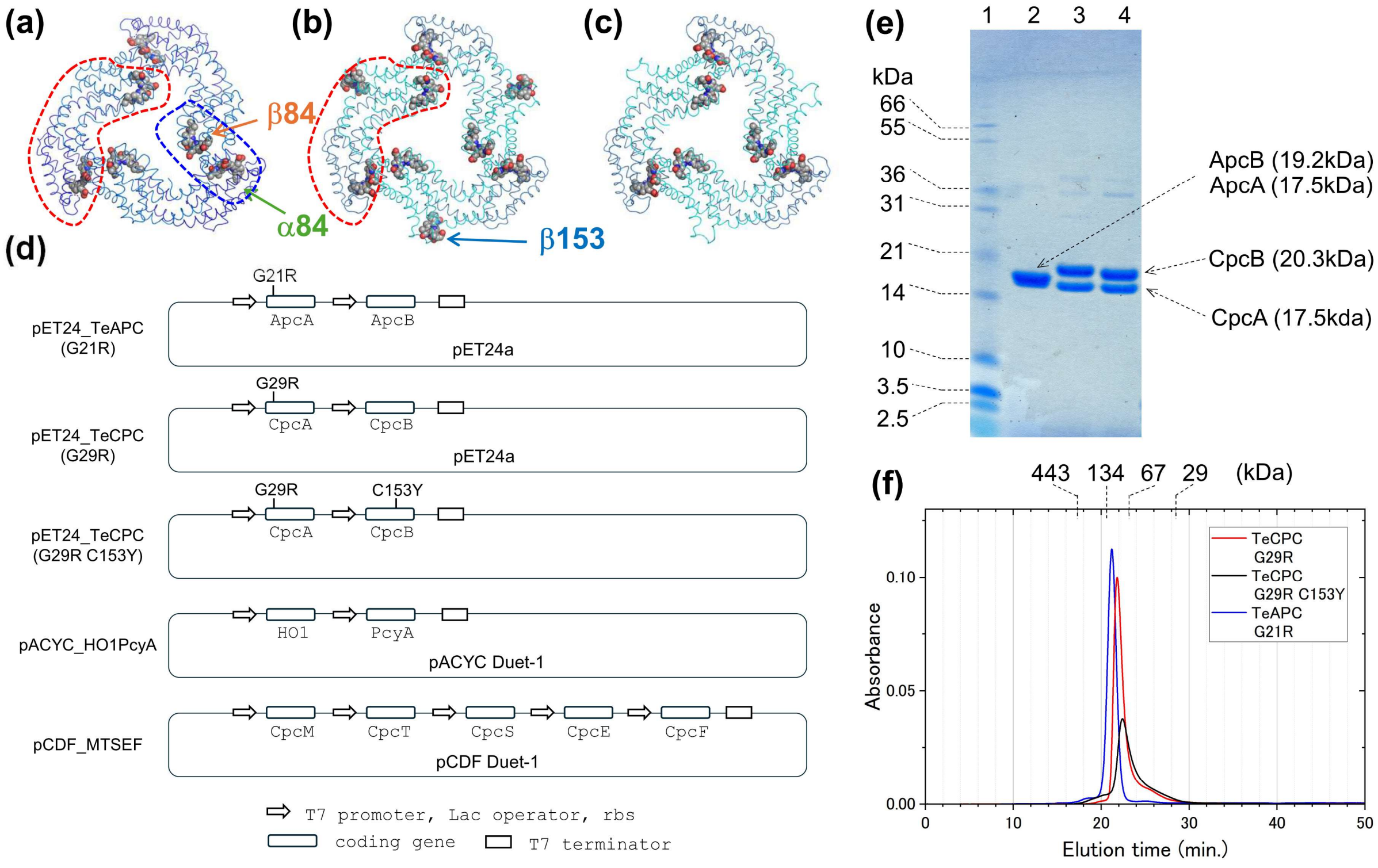

Fig. 1. Sample production and characterization. (a-c) Structures of the wild-type APC trimer, wild-type CPC trimer, and mutant CPC trimer (PCB-deficient CPC at the β153 site), respectively. Phycocyanobilin (PCB) pigments at positions α84, β84, and β153 are shown in a van der Waals representation. The red dashed line indicates a monomer boundary, and the blue dashed line encloses a pair of pigments forming excitonic states. The figures were made using the software "Pymol" and PDB entries 3DBJ and 3L0F, corresponding to wild-type APC and CPC, respectively. (d) Expression plasmids used for protein synthesis. Three plasmids, pET24_TeAPC G21R, pET24_TeCPC G29R and pET24_TeCPC G21R C153Y, were constructed to produce the proteins shown in (a), (b), and (c), respectively. Coexpression plasmids, pACYC_HO1PcyA and pCDF_MTSEF (for APC/CPC modification by encoded enzymes in *E. coli*), were also used.[27] (e) SDS-PAGE analysis. Lane 1: Mark12 Unstained Standard; Lanes 2-4: TeAPC G21R, TeCPC TeG29R, and TeCPC G29R C153Y, respectively. (f) Size-exclusion chromatography profiles.

Arrows indicate elution times of standard proteins: apoferritin (443 kDa, 17.32 min), dimer of bovine serum albumin (BSA) (134 kDa, 20.60 min), BSA (67 kDa, 23.14 min), and carbonic anhydrase (29 kDa, 28.50 min).

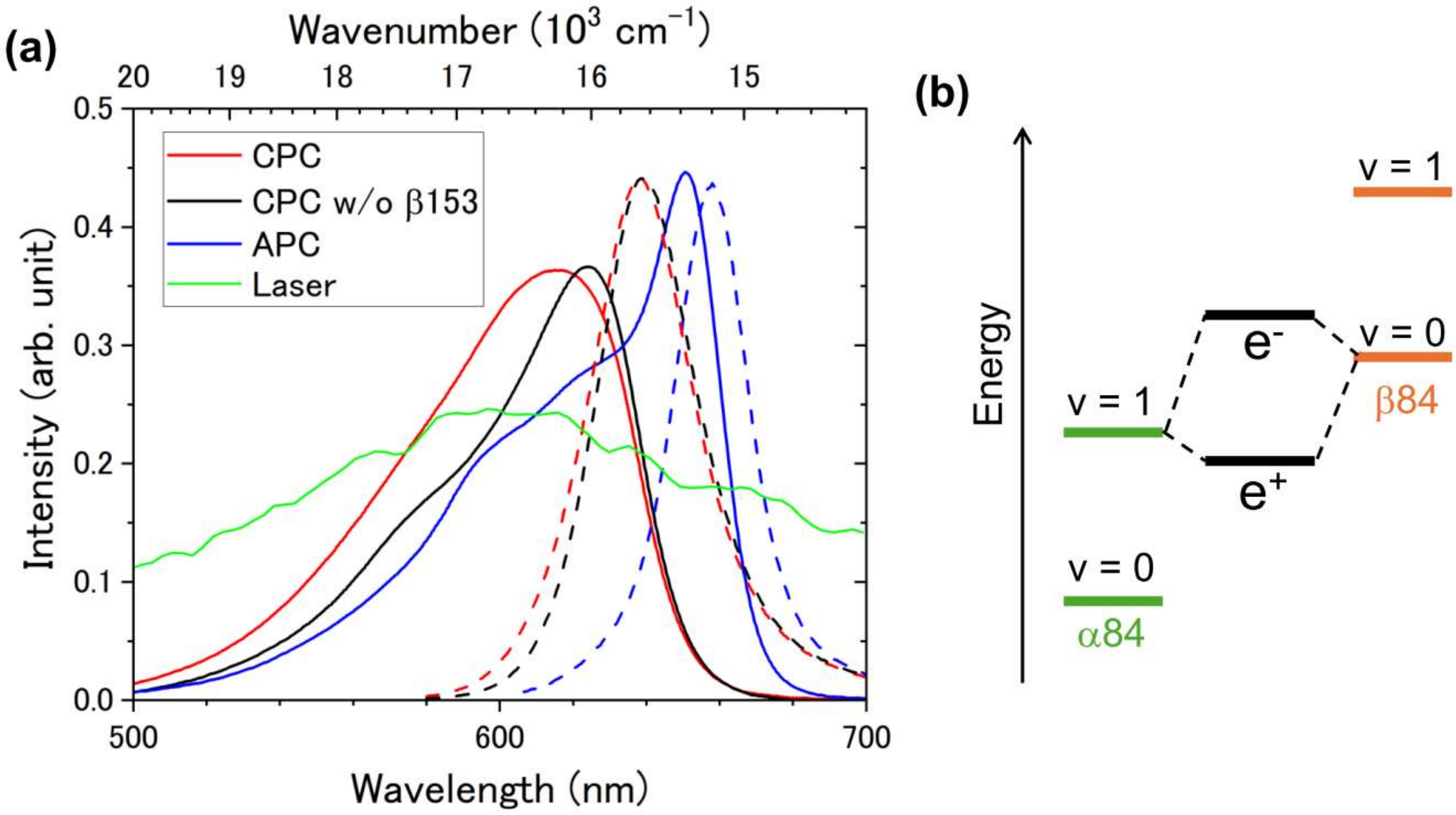

Fig. 2. (a) Absorption and fluorescence spectra. The solid lines show the absorption spectra of wild-type APC (blue), wild-type CPC (red), and mutant CPC (black). The fluorescence spectra are shown as dashed lines which were measured using excitation wavelengths of 600 nm for APC and 570 nm for CPC and the mutant CPC. The thin green line represents the laser spectrum, which spans the full absorption and fluorescence ranges of all proteins. (b) Energy-level diagram illustrating vibronic coupling model.[20]

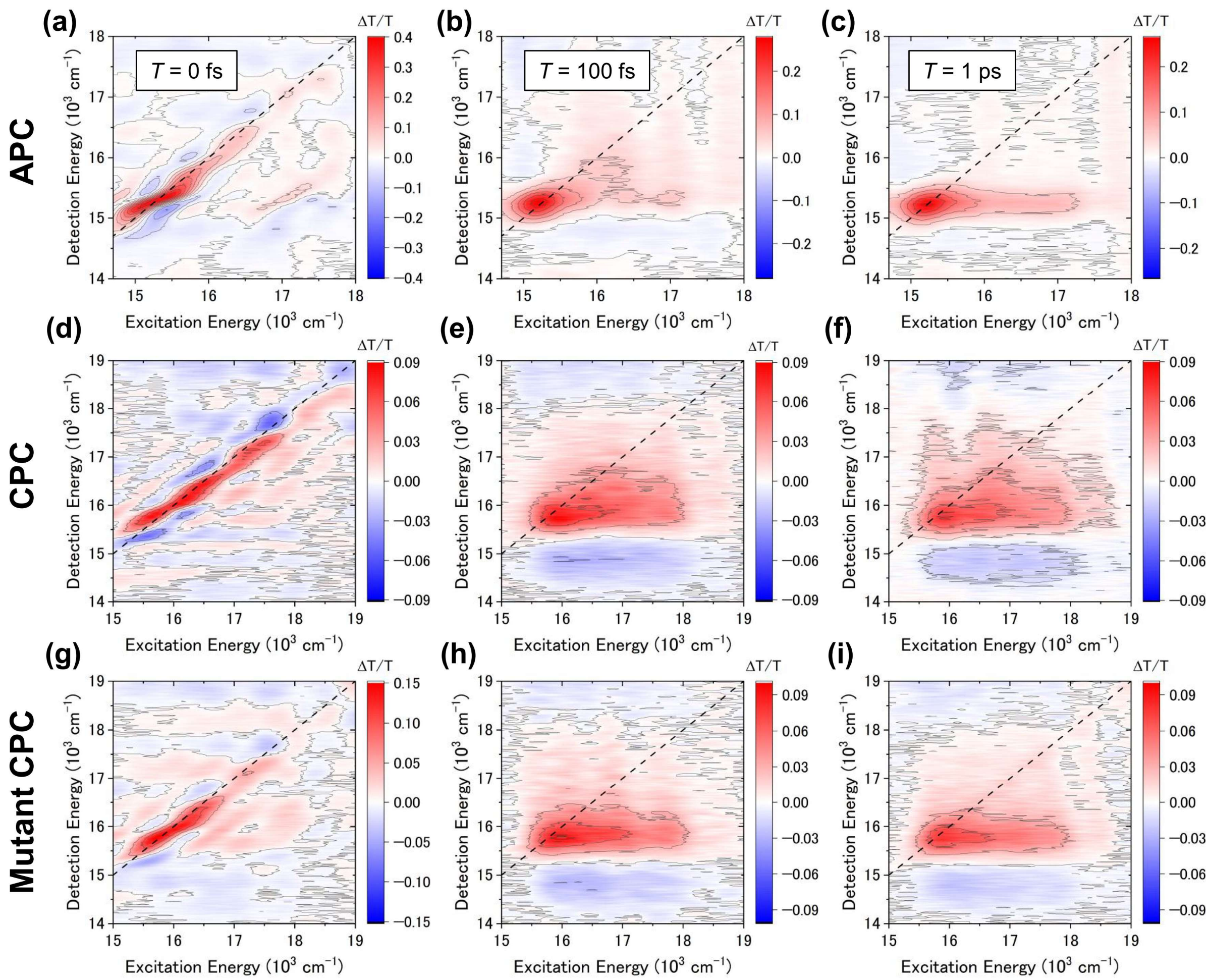

Fig. 3. A series of 2D-E spectra, $S(E_{\mathrm{exc}}, E_{\mathrm{det}}; T)$. Panels (a-c) show the 2D-E spectra of APC at pump-probe delay time of (a) 0 fs, (b) 100 fs, and (c) 1 ps. Panels (d-f) show the corresponding spectra of CPC at (d) 0 fs, (e) 100 fs, and (f) 1 ps. Panels (g-i) show the spectra of the mutant CPC at (g) 0 fs, (h) 100 fs, and (i) 1 ps. The horizontal and vertical axes represent the excitation ($E_{\mathrm{exc}}$) and detection ($E_{\mathrm{det}}$) energies, respectively. Positive amplitudes correspond to GSB or SE, whereas negative amplitudes correspond to ESA. The dashed line indicates the diagonal.

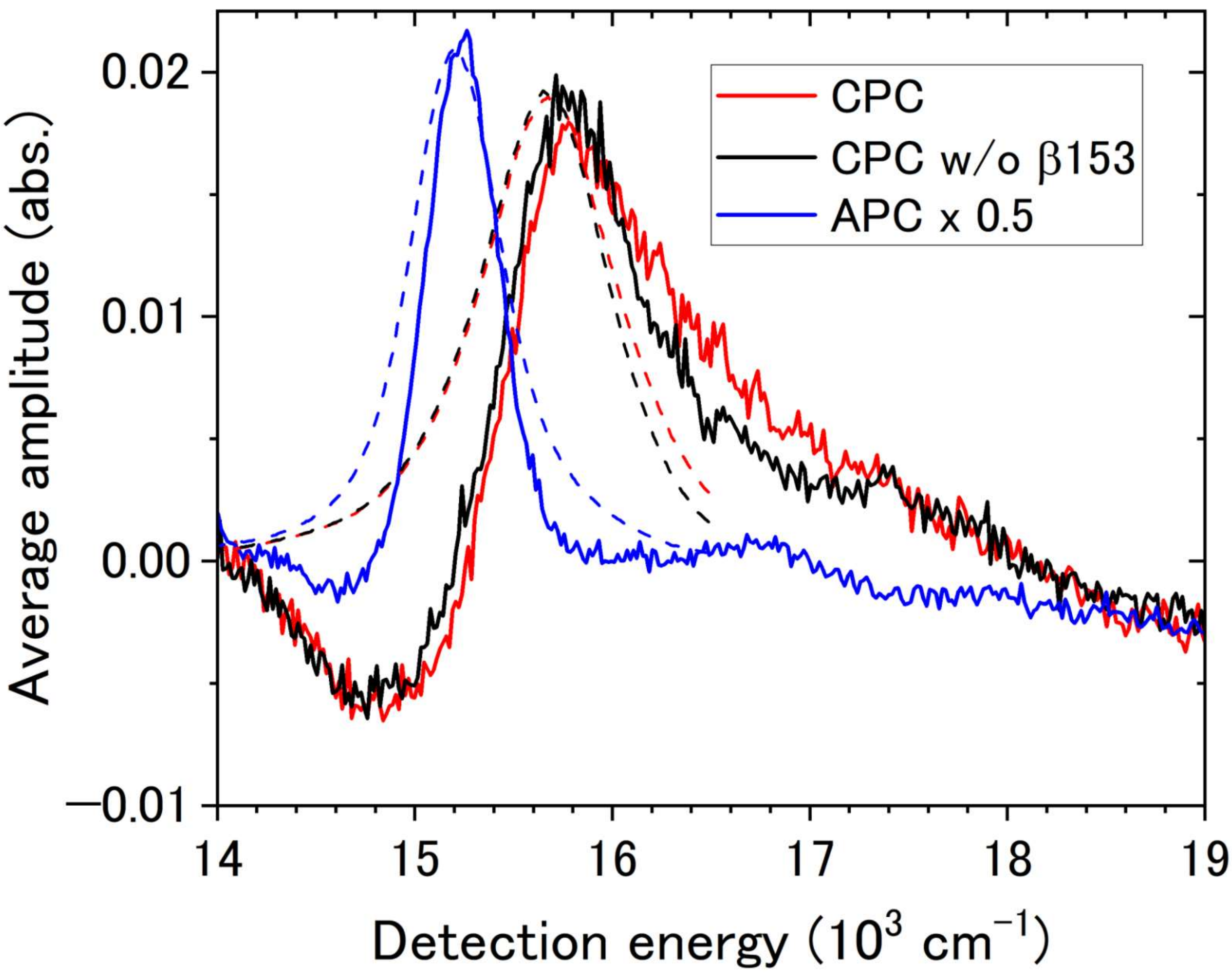

Fig. 4. Comparison of transient spectra (thick solid lines) at a long delay time ($T$ = 2 ps) with fluorescence spectra (thin dashed lines). The transient spectra were obtained by integrating the 2D-E spectra along the excitation energy axis. Red, black and blue lines represent the spectra of wild-type CPC, mutant CPC, and wild-type APC, respectively.

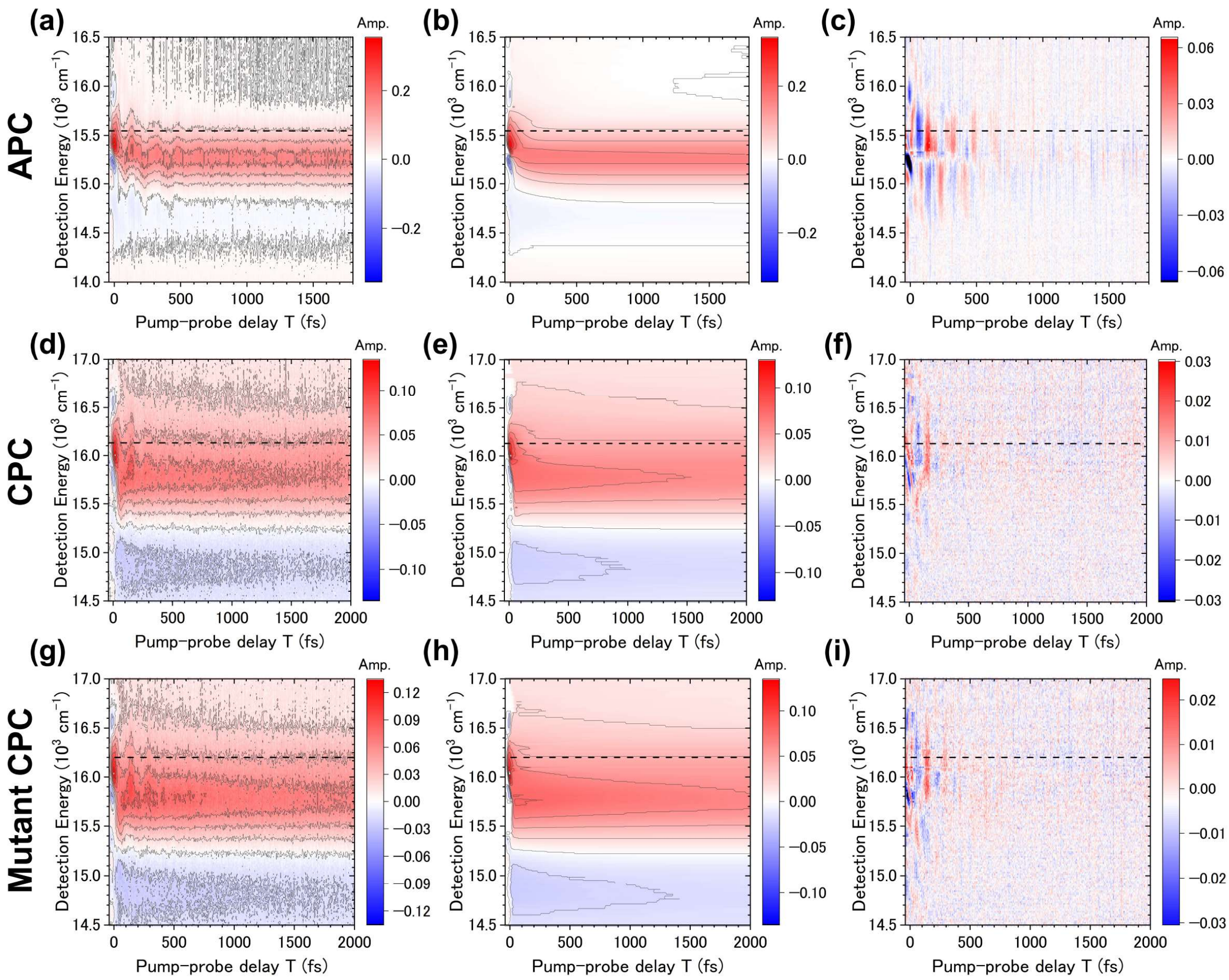

Fig. 5. Transient spectra, $S(E_{\mathrm{det}}, T; E_{\mathrm{exc}})$, following excitation to the lowest electronic excited state. Panel (a) shows the transient spectrum of APC obtained by slicing the three-dimensional data set $S(E_{\mathrm{exc}}, E_{\mathrm{det}}, T)$ at an excitation energy of $E_{\mathrm{exc}}$ = 15,550 cm$^{-1}$. This spectrum is further decomposed into the (b) kinetic component and (c) coherent dynamics. Panels (d) and (g) show the transient spectra of wild-type and mutant CPCs at excitation energies of $E_{\mathrm{exc}}$ = 16,130 cm$^{-1}$ and 16,200 cm$^{-1}$, respectively. Panels (e) and (f) present the kinetic and coherent components of wild-type CPC, respectively, while panels (h) and (i) show the corresponding components of mutant CPC. Horizontal dashed lines indicate the excitation energies.

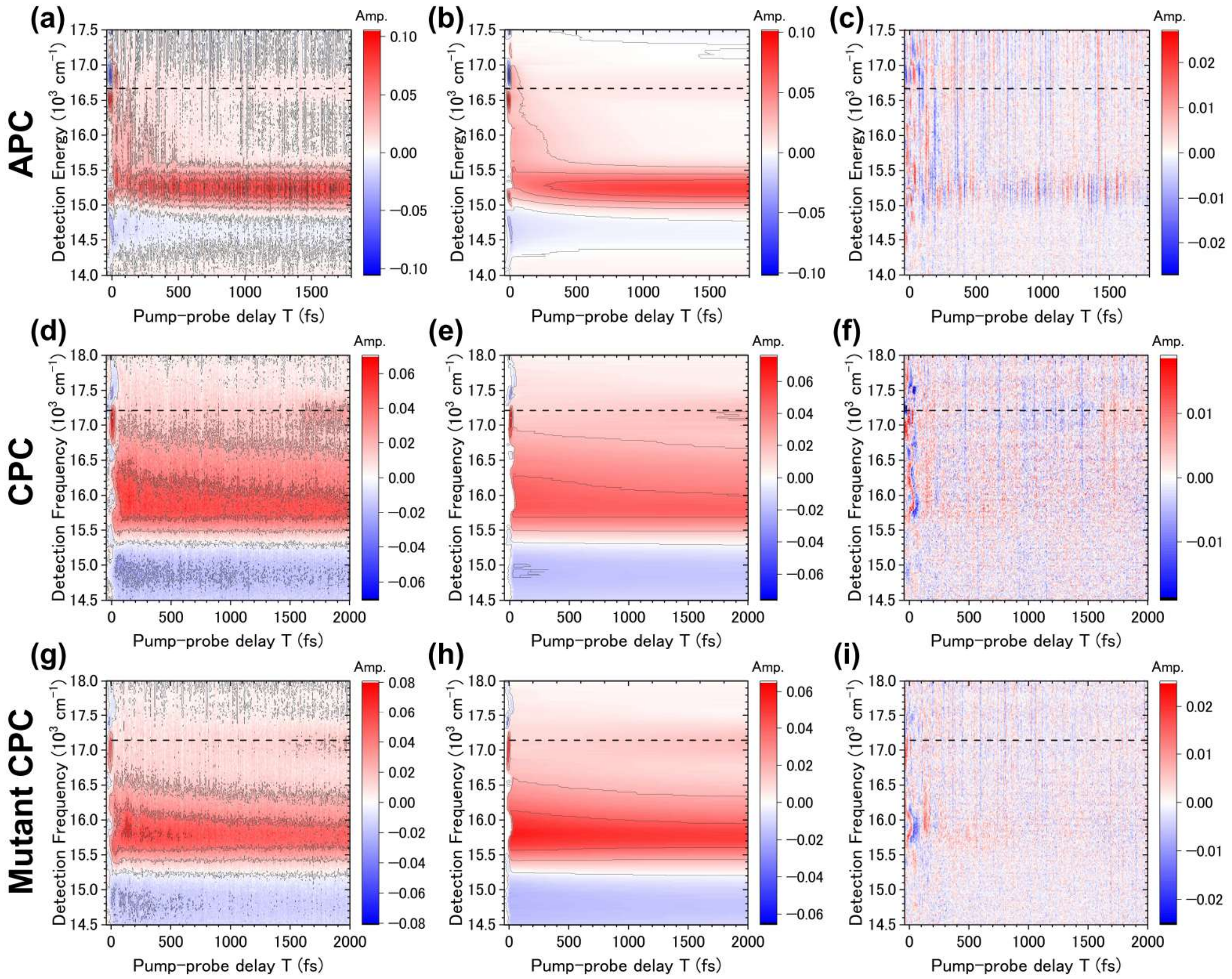

Fig. 6. Transient spectra, $S(E_{\mathrm{det}}, T; E_{\mathrm{exc}})$, following excitation to the higher excitonic state. Panel (a) shows the transient spectrum of APC obtained by slicing the three-dimensional data set $S(E_{\mathrm{exc}}, E_{\mathrm{det}}, T)$ at an excitation energy of $E_{\mathrm{exc}}$ = 16,670 cm$^{-1}$. This spectrum is further decomposed into the (b) kinetic component and (c) coherent dynamics. Panels (d) and (g) show the transient spectra of wild-type and mutant CPCs at excitation energies of $E_{\mathrm{exc}}$ = 17,210 cm$^{-1}$ and 17,140 cm$^{-1}$, respectively. Panels (e) and (f) present the kinetic and coherent components of wild-type CPC, respectively, while panels (h) and (i) show the corresponding components of mutant CPC. Horizontal dashed lines indicate the excitation energies.

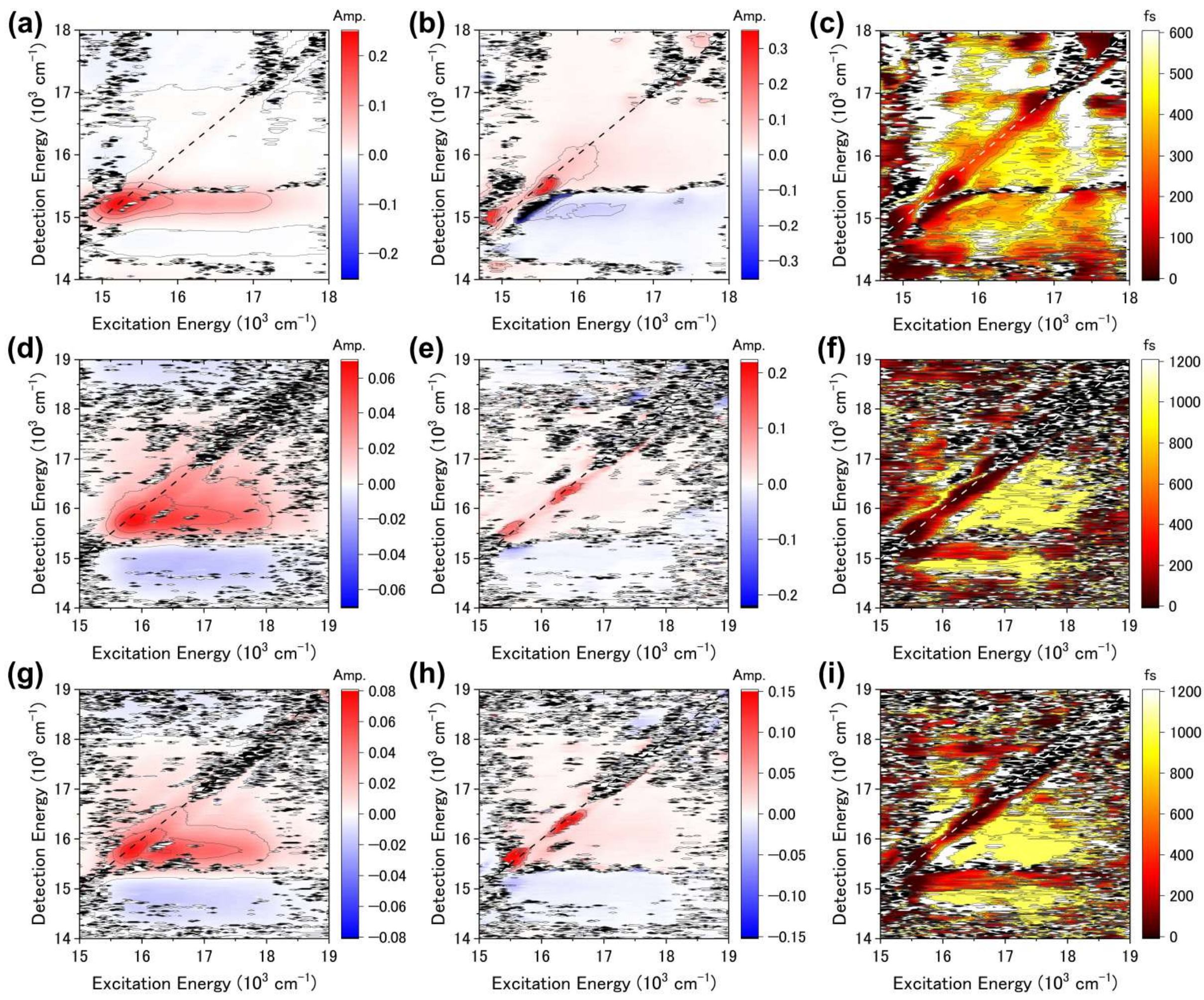

Fig. 7. Two-dimensional parameter maps. Panels (a-c) show the maps of parameters $A_0$, $A_1$, and $\tau_1$ for wild-type APC derived from least-squares fitting, as described in Supplementary Section 4. Panels (d-f) present the corresponding $A_0$, $A_1$, and $\tau_1$ maps for wild-type CPC. Panels (g-i) present the $A_0$, $A_1$, and $\tau_1$ maps for the β153-deficient CPC mutant. The dashed line indicates the diagonal.

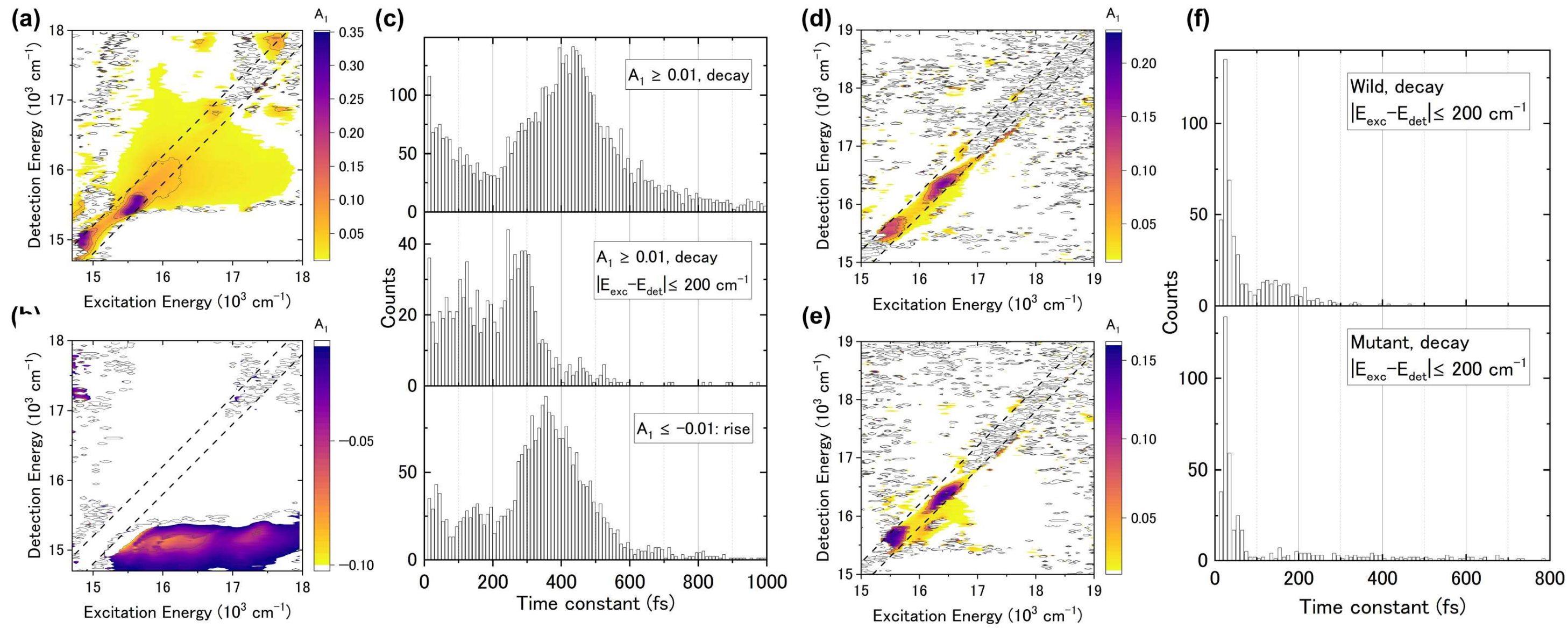

Fig. 8. Distributions of the time constants $\tau_1$ obtained at individual $(E_{\text{exc}}, E_{\text{det}})$ grid points. (a, b) The 2D-DASs ($A_1$ map) of APC, highlighting only (a) positive and (b) negative components, respectively. A lower threshold on the absolute amplitude of 0.01 was applied to exclude regions exhibiting negligible population decay or rise. (c) Histograms of $\tau_1$ for APC. The upper and lower panels show the distributions for regions with positive and negative amplitudes in 2D-DAS, respectively, while the middle panel presents the histogram for the region bounded by dashed lines in panel (a). (d, e) 2D-DASs of (d) wild-type and (e) mutant CPCs, shown with a lower amplitude threshold of 0.015. (f) Histograms of $\tau_1$ for wild-type and mutant CPCs, shown in the upper and lower panels, respectively.

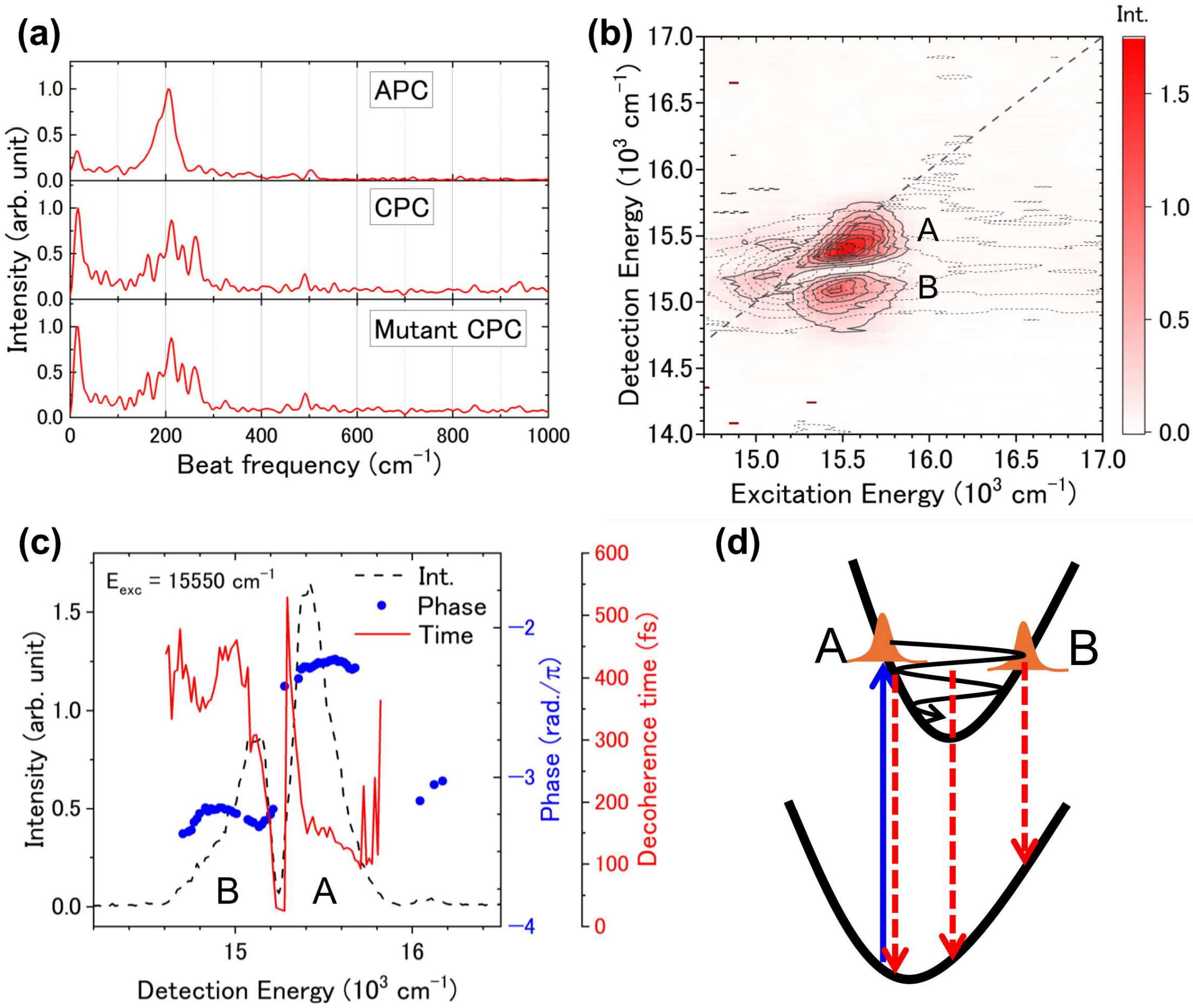

Fig. 9. (a) Beat-frequency spectra of wild-type APC, wild-type CPC, and mutant CPC measured at excitation energies of $E_{\mathrm{exc}}$ = 15,550 cm$^{-1}$, 16,130 cm$^{-1}$ and 16,200 cm$^{-1}$, respectively. The spectra were obtained by averaging over detection energy range of $E_{\mathrm{det}}$ = 15,000 ~ 15,650 cm$^{-1}$ for APC and $E_{\mathrm{det}}$ = 15,750 ~ 16,500 cm$^{-1}$ for CPCs. (b) Beat-frequency-resolved 2D-E spectra of APC averaged over beat frequencies of $\nu_{\mathrm{beat}}$ = 202 ± 46 cm$^{-1}$. The dashed line indicates the diagonal. The 2D-E spectrum at $T$ = 1 ps is overlaid as a thin dashed contour map. (c) Beat intensity (black dashed line) and phase (blue circles) spectra of APC at an excitation energy $E_{\mathrm{exc}}$ = 15,550 cm$^{-1}$. The decoherence time as a function of detection energy is also overlaid as a red solid line. (d) Schematic illustration of a vibrational wave packet on the electronic excited state.

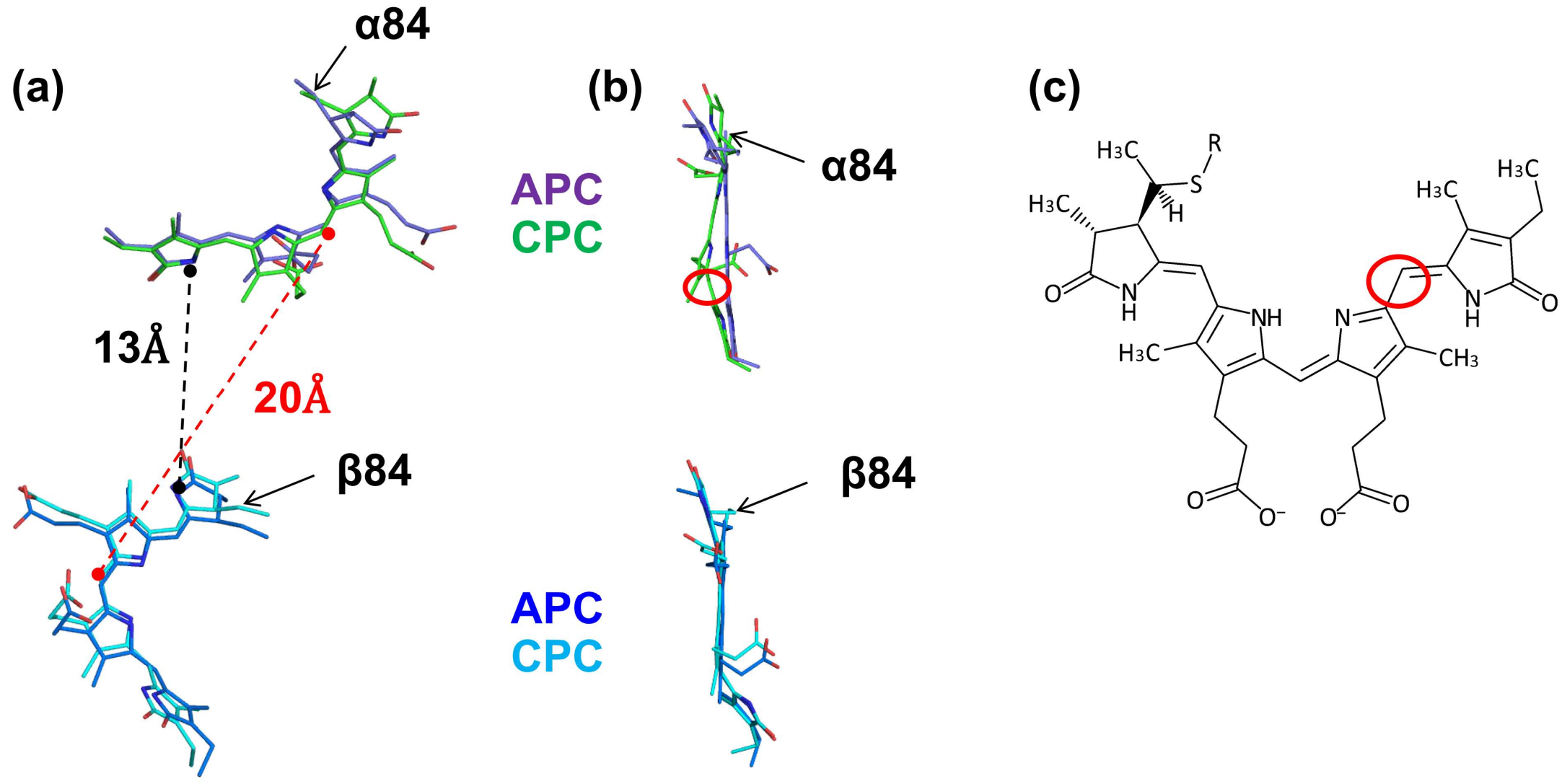

Fig. 10. Structures of phycocyanobilin (PCB) extracted from PDB entries 3DBJ and 3L0F, corresponding to wild-type APC and CPC, respectively. (a) Top views of PCB in the α-chain (purple) and β-chain (blue) of APC, and in the α-chain (green) and β-chain (cyan) of CPC. Arrows indicate the binding sites to the α84 and β84 cysteine residues of proteins. (b) Side views of PCB. Relative positioning is not preserved. (c) Structural formula of 3Z-Phycocyanobilin. The red circle highlights the C−C bond associated with the structural difference in the α84 PCB between APC and CPC.

# Supplemental Issues

## Roles of Individual Pigments in Ultrafast Excitation Dynamics of Light-Harvesting Phycobiliproteins Revealed by Recombinant Techniques and Two-Dimensional Electronic Spectroscopy

**Masaaki Tsubouchi,[1,2,a)] Takatoshi Fujita,[2] Motoyasu Adachi,[2] and Ryuji Itakura[1]**

**AFFILIATIONS**

[1]Kansai Institute for Photon Science (KPSI), National Institutes for Quantum Science and Technology (QST), 8-1-7 Umemidai, Kizugawa, Kyoto 619-0215, Japan

[2]Institute for Quantum Life Science, National Institutes for Quantum Science and Technology (QST), 4-9-1 Anagawa, Inage, Chiba 263-8555, Japan

[a)]Author to whom correspondence should be addressed: tsubouchi.masaaki@qst.go.jp

## 1. Sample preparation

The recombinant proteins allophycocyanin (APC) and C-phycocyanin (CPC) from *Thermosynechococcus elongatus* BP-1 were independently produced by an *Escherichia coli* (*E. coli*) expression system. Supplementary Table S1 lists the expression plasmids used in this study. The DNA and encoded amino acid sequences of the proteins were deposited into the DNA Data Bank of Japan. The sequences were designed based on three-dimensional structure [1-6]. APC and CPC proteins typically assemble into hexameric structures of heterodimers of α- and β-chains within the phycobilisome supercomplex. To facilitate spectroscopic and biochemical experiments, we introduced mutations at the intertrimeric interface to promote trimer formation.

The α-chain of CPC and α- and β-chains of APC link one phycocyanobilin (PCB) as a covalently bound pigment in native phycobilisomes, whereas the β-chains of CPC link two PCBs. Two plasmids, pACYC_HO1PcyA and pCDF_MTSEF, were used for coexpression to enable pigment synthesis and protein modification, respectively. Specifically, Gly21 and Gly29 in the α-chains of APC and CPC were substituted with arginine residues to introduce steric hinderance in hexamer formation. Furthermore, to match the pigment configuration of CPC to that of APC, we generated a mutant CPC lacking the β153 PCB pigment by substituting Cys153 with a tyrosine residue. The C153Y mutation in pET_TeCPC was introduced by the PCR method using PrimeSTAR Max DNA Polymerase (Takara, Japan) and DNA primers of TeCPC_C153Y_F (5’-GGCGATTACAGCGCGCTGATGAGCGAAATC-3’) and TeCPC_C153Y_R (5’-CGCGCTGTAATCGCCCGGGGTGATACCGTT-3’). Plasmids were prepared using QIAprep Spin Miniprep Kit (Qiagen). The DNA sequence coding for the proteins was confirmed by DNA sequencing on the SeqStudio™ Genetic Analyzer (Thermo Fisher Scientific). In the following, unless otherwise specified, APC and CPC refer to the APC trimer and the CPC trimer, respectively. All DNAs were chemically synthesized and purchased from Genewiz (Azenta Life Sciences, Japan), and were optimized for *E. coli*. The expression plasmids were transformed into *E. coli* JM109(DE3) (Promega).

**Table S1** | List of expression plasmids used in this study.

| Name of plasmid | Original plasmid | Coded proteins | Organisms* | Accession number (reference sequence) | Accession number deposited |
|---|---|---|---|---|---|
| pET_TeAPC (G21R) [APC] | pET24a (Novagen) | ApcA | BP-1 | NCBI: WP_011056801 | LC912691 |
| | | ApcB | BP-1 | NCBI: WP_011056800 | |
| pET_TeCPC (G29R) [CPC] | pET24a (Novagen) | CpcA | BP-1 | NCBI: WP_011057793 | LC912692 |
| | | CpcB | BP-1 | NCBI: WP_011057792 | |
| pET_TeCPC (G29R C153Y) [CPC] | pET24a (Novagen) | CpcA | BP-1 | NCBI: WP_011057793 | Subcloned |
| | | CpcB | BP-1 | NCBI: WP_011057792 | |
| pACYC_ HO1PcyA [7] | pACYCDuet-1 (Novagen) | HO1 | 6803 | NCBI: WP_010871494 | LC853284 |
| | | PcyA | 7120 | NCBI: WP_010997850 | |
| pCDF_ MTSEF [7] | pCDFDuet-1 (Novagen) | CpcM | BP-1 | NCBI: WP_011057783 | LC853283 |
| | | CpcT | 7120 | NCBI: WP_010999463 | |
| | | CpcS | 7120 | NCBI: WP_010994793 | |
| | | CpcE | 7120 | NCBI: WP_010994708 | |
| | | CpcF | 7120 | NCBI: WP_010994709 | |

*BP-1: *Thermosynechococcus elongatus* BP-1 (*Thermosynechococcus vestitus* BP-1)

6803: *Synechocystis* sp. PCC 6803

7120: *Nostoc* sp. 7120 (*Anabaena* sp.)

## 2. Sample characterization

### SDS-PAGE analysis

NuPAGE Bis-Tris Precast Gel (Thermo Fisher Scientific) and 2-(N-morpholino)ethanesulfonic acid buffer were used for electrophoretic analysis. Electrophoresis was performed at a constant voltage of 100 V for 80 min, and the gels were subsequently stained with EzStain AQua (ATTO).

### Mass spectrometry analysis

Purified protein samples were analyzed using an ultra-high-performance liquid chromatography system coupled to a quadruple time-of-flight mass spectrometer (H-Class and Xevo G2-XS; Milford, MA, USA, Waters). Intact proteins were desalted via chromatographic separation on a MassPREP Micro Desalting Column (Waters) using a gradient from water to acetonitrile containing 0.1% formic acid. Leucine enkephalin and NaI were employed for reference ion correlation and calibration, respectively. Raw data generated by MassLynx were processed using UNIFI version 1.9.

### Gel filtration analysis

A Superdex 200 10/300GL gel filtration column was equilibrated with Buffer T. A 0.03 mL sample at a concentration of 5 μM was injected and eluted at a flow rate of 0.5 mL/min. Protein elution was monitored by measuring absorbance at 280 nm.

In the size-exclusion chromatogram, detection is achieved with an absorbance of approximately 0.1, which corresponds to a concentration over 100-fold lower than that used in the 2D-ES measurements. Even under such diluted conditions, a peak is observed at the position corresponding to trimer formation. In general, oligomer formation is promoted at higher concentrations. Regarding glycerol, it is commonly added during refolding because it suppresses protein aggregation; however, at a concentration of 10%, it is not strong enough to disrupt the quaternary structure of proteins. Although a weak shoulder is observed under highly diluted condition as shown in Fig. 1(f) in main text, its contribution is minor compared to the dominant trimer peak. Therefore, under the 2D-ES measurement conditions, the protein is predominantly in the oligomeric (trimeric) state.

## 3. Analysis of absorption and fluorescence spectra

The absorption spectra of all three phycobiliproteins are well fitted by the sum of three Gaussian functions as shown in Supplementary Fig. S1(a). The fits using only two Gaussian components result in large residuals. It is possible that this discrepancy arises from the fact that, instead of employing a spectral shape that accounts for the anharmonic potential energy surface, a Gaussian line shape is assumed. Nevertheless, in the present study, we proceed with the analysis by assuming three spectral components. The three components are referred to as peaks A, B, and C from the long-wavelength side, and the corresponding peak energies and spectral widths are summarized in Table 1 in the main text.

The fluorescence spectra are fitted by the sum of two Gaussian functions as shown in Supplementary Fig. S1(b). However, in principle, the analysis of the fluorescence spectra should also be performed based on an anharmonic potential.

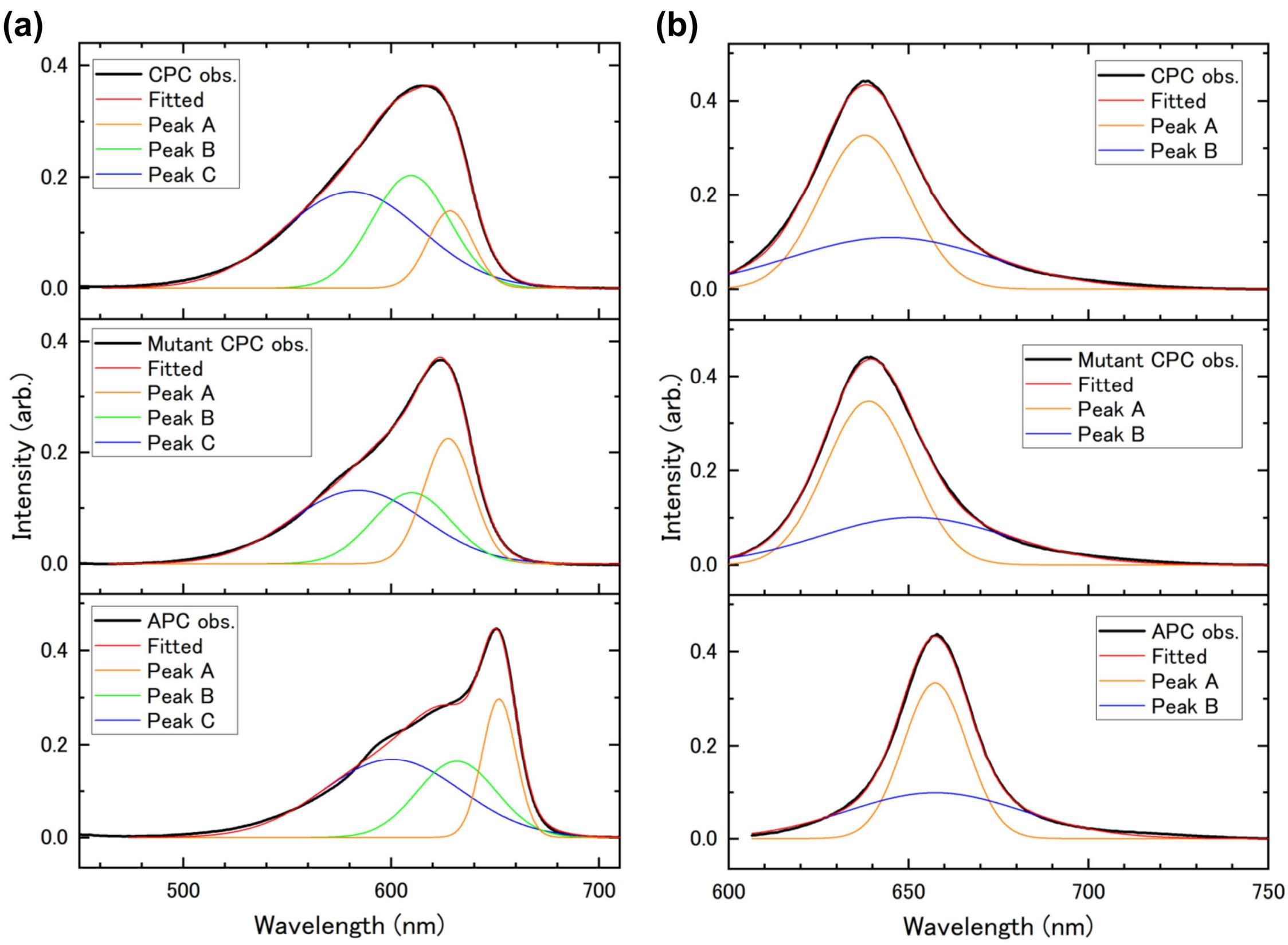

Figure S1. Nonlinear least-squares fitting of experimental (a) absorption and (b) fluorescence spectra using sum of three and two Gaussian functions, respectively. The upper, middle, and lower panels show the results for wild-type CPC, mutant CPC, and wild-type APC, respectively. The black and red lines represent the experimental spectra and fitted results, respectively.

## 4. Analysis of two-dimensional electronic spectra

To understand the kinetics of the photoexcited molecules, global analysis has been widely applied to the transient spectra. However, when coherent features overlap with kinetic components, applying Global analysis becomes difficult, because the experimental data cannot be accurately reconstructed using a small number of principal components obtained through singular value decomposition. Therefore, the time profiles obtained at all pairs of excitation and detection energies are fitted with a single exponential function,

$$S(T; E_{\mathrm{exc}}, E_{\mathrm{det}}) = A_0 + A_1 \exp(-T/\tau_1), \tag{S1}$$

where $E_{\mathrm{exc}}$ and $E_{\mathrm{det}}$ are the excitation and detection energies, respectively. $T$ is the delay time after photoexcitation. $\tau_1$ is the time constant of the single exponential decay. $A_0$ and $A_1$ are the amplitudes of the constant and the decay components, respectively. These three parameters are functions of $E_{\mathrm{exc}}$ and $E_{\mathrm{det}}$.

In the fitting procedure, we also incorporate the instrumental response function (IRF) and account for the coherent artifact (CA) arising from non-resonant interactions between the pump and probe pulses near zero delay, as discussed in Refs [8, 9]. IRF is defined in Gaussian form as

$$IRF(T) = \frac{1}{d\sqrt{2\pi}} \exp\left[-\frac{(T-T_0)^2}{2d^2}\right], \tag{S2}$$

where $T_0$ is the delay time corresponding to the IRF maximum, and $d = \Delta/\left(2\sqrt{2\ln 2}\right)$ with $\Delta$ representing the full width at half maximum (FWHM) of the IRF. This IRF function is convolved with the exponential decay function Eq. (S1), and we obtain,

$$\begin{aligned} S(T;\, E_{\mathrm{exc}}, E_{\mathrm{det}}) * IRF(T) &= \frac{A_0}{2}\left[1 + \mathrm{erf}\left(\frac{T-T_0}{d\sqrt{2}}\right)\right] \\ &+ \frac{A_1}{2}\left[1 + \mathrm{erf}\left(\frac{T - T_0 - \frac{d^2}{\tau_1}}{d\sqrt{2}}\right)\right] \exp\left(-\frac{T}{\tau_1}\right) \exp\left(\frac{T_0 + \frac{d^2}{2\tau_1}}{\tau_1}\right). \end{aligned} \tag{S3}$$

The CA component is described by the sum of the zeroth, first, and second derivatives of the IRF. The time profiles obtained at all pairs of excitation and detection energies are fitted with $S(T;\, E_{\mathrm{exc}}, E_{\mathrm{det}}) * IRF(T) + CA(T)$.

Supplementary Fig. S2 (a) shows the fitting results of the time profiles of wild-type APC using the model function described in this section. The profiles are obtained at an excitation energy of $E_{\mathrm{exc}} = 15{,}550\ \mathrm{cm}^{-1}$. The kinetic component and coherent dynamics are then separated based on the fitted profiles and their residuals, as shown in Supplementary Fig. S2 (b).

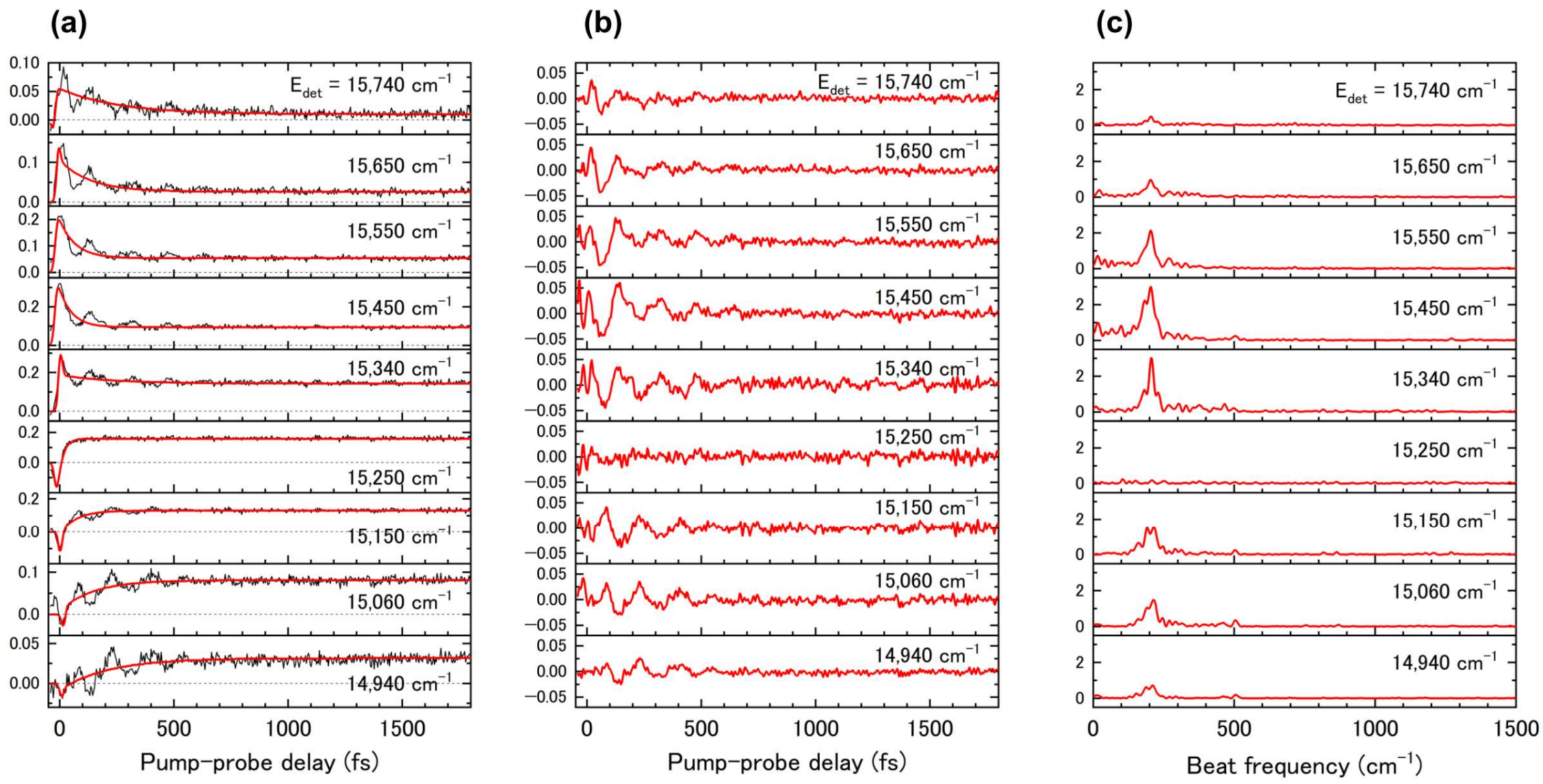

Figure S2. (a) Fitting results of the time profiles using the model function $S(T; E_{\mathrm{exc}}, E_{\mathrm{det}}) * IRF(T) + CA(T)$. Black and red lines represent the observed and fitted time profiles, respectively, for wild-type APC at an excitation energy of $E_{\mathrm{exc}}$ = 15,550 cm$^{-1}$. (b) Residual time profiles obtained by subtracting the fitted profiles from the observed ones. (c) Fourier-transformed spectra of the residual time profiles.

## 5. Analysis of vibrational wave packet

Supplementary Fig. 2(c) presents the Fourier-transformed spectra of the residual time profiles. A distinct peak is observed at the beat frequency $\nu_{\text{beat}} = 202$ cm$^{-1}$. At the nodal line ($E_{\text{det}}$ = 15,250 cm$^{-1}$) of the two-dimensional map shown in Fig. 9(b) of the main text, the amplitude of the beat spectrum decreases to zero.

To evaluate the decoherence time of the vibrational coherence at the beat frequency of 202 cm$^{-1}$, the observed time profiles are fitted to a single exponential decay function combined with a damped oscillation as follows. For this purpose, Eq. (S1) is modified as follows:

$$S_{\text{OSC}}(T;\, E_{\text{exc}}, E_{\text{det}}) = S(T; E_{\text{exc}}, E_{\text{det}}) + A_{\text{OSC}} \cos[2\pi\nu_{\text{beat}}(T - T_{\text{OSC}})] \exp(-T/\tau_{\text{OSC}}), \qquad \text{(S4)}$$

where $A_{\text{OSC}}$, $T_{\text{OSC}}$, and $\tau_{\text{OSC}}$ represent the amplitude, phase delay, and decoherence time of the beat signal, respectively.

Supplementary Fig. S3 shows the fitting results of the time profiles of wild-type APC using the model function given in Eq. (S4). The profiles are obtained at an excitation energy of $E_{\text{exc}}$ = 15,550 cm$^{-1}$. The decoherence times determined by this fitting are presented in Fig. 9(c) of the main text as the red line.

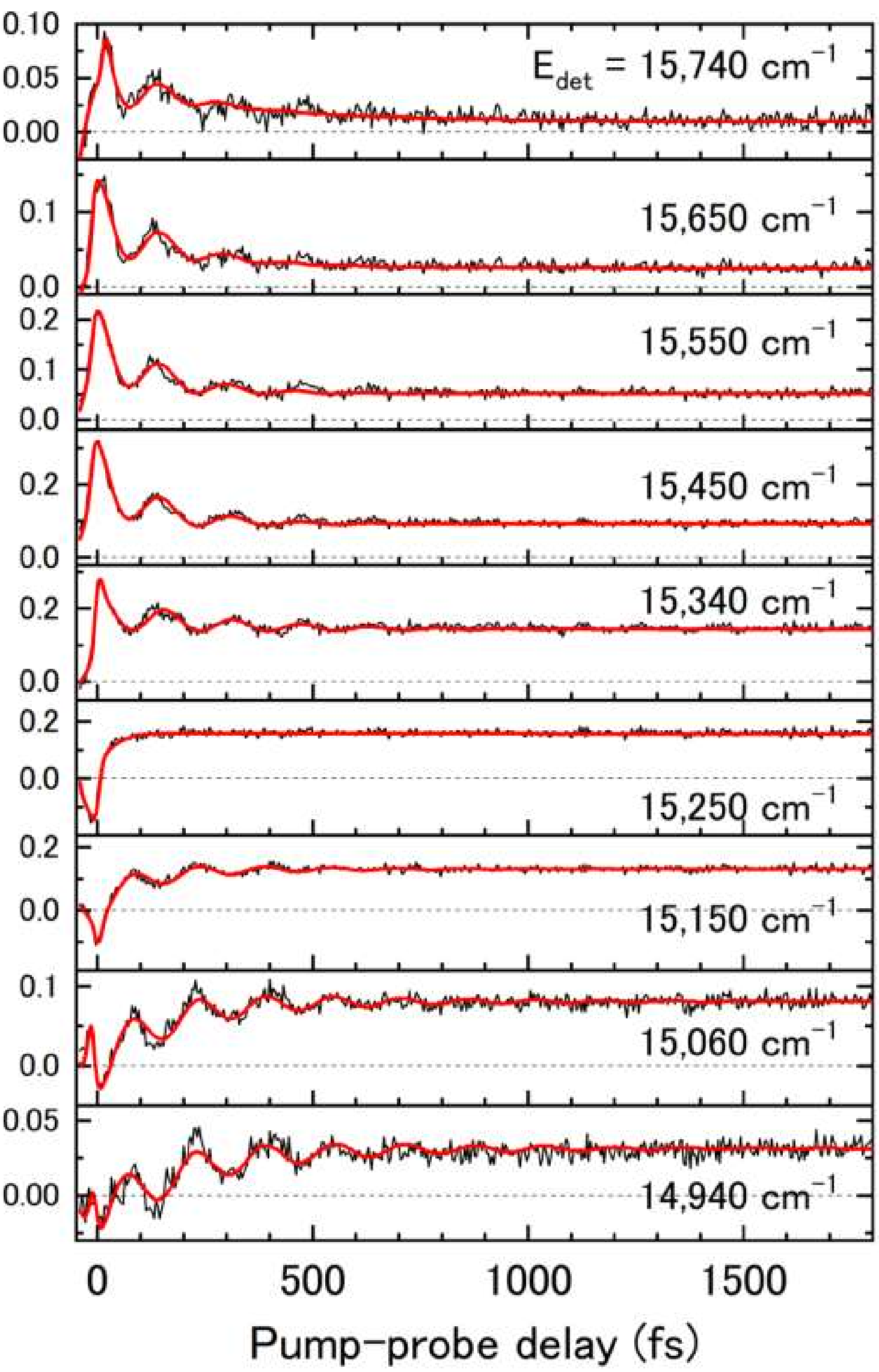

Figure S3. Fitting results of the time profiles using the model function given in Eq. (S4). Black and red lines represent the observed and fitted time profiles, respectively, for wild-type APC at an excitation energy of $E_{exc}$ = 15,550 cm$^{-1}$.